\documentclass[11pt,a4paper]{article}
\usepackage{jheppub,amsmath,  amssymb,slashed,url,bm,textgreek,upgreek}
\usepackage{graphicx}
\usepackage{epstopdf}
\def\t{{ \sf t}} 

\def\tt{{\mathfrak t}}

\def\SO{{\mathrm{SO}}}

\def\a{a}

\def\la{\langle}
\def\ra{\rangle}
\def\Tr{{\mathrm{Tr}}}
\def\max{{\mathrm{max}}}

\def\dS{{\mathrm{dS}}}
\def\add{{\mathrm{add}}}

\def\sD{{\sf D}}

\def\RR{{\mathcal R}}

\def\UU{{\mathrm{U}}}

\def\be{\begin{equation}}
\def\ee{\end{equation}}

\def\Im{{\mathrm{Im}}}

\def\h{\widehat}

\def\V{{\mathcal V}}
\def\O{{\mathcal O}}

\def\A{{\mathcal A}}
\def\d{{\mathrm d}}

\def\b{\overline}
\def\R{{\mathbb R}}

\def\U{{\mathcal U}}
\def\E{{\mathcal E}}

\def\[{\bigl [}

\def\]{\bigr ]}

\def\tr{{\mathrm {tr}}}
\def\Z{{\mathbb Z}}

\def\CC{{\mathcal C}}

\def\r{{\mathrm{tr}}}

\def\t{\widetilde }
\def\h{\widehat}

\def\V{{\mathcal V}}

\def\B{{\mathcal B}}

\def\l{\langle}

\def\r{\rangle}

\def\H{{\mathcal H}}

\def\bar{\overline}
\def\l{\langle}
\def\r{\rangle}
\def\a{{\sf a}}
\def\b{{\sf b}}

\def\i{{\mathrm i}}

 \font\teneurm=eurm10 \font\seveneurm=eurm7  \font\fiveeurm=eurm5
 \newfam\eurmfam
 \textfont\eurmfam=\teneurm \scriptfont\eurmfam=\seveneurm
 \scriptscriptfont\eurmfam=\fiveeurm
 
 \font\teneusm=eusm10 \font\seveneusm=eusm7 \font\fiveeusm=eusm5
 \newfam\eusmfam
 \textfont\eusmfam=\teneusm \scriptfont\eusmfam=\seveneusm
 \scriptscriptfont\eusmfam=\fiveeusm
 
 \font\tencmmib=cmmib10 \skewchar\tencmmib='177
 \font\sevencmmib=cmmib7 \skewchar\sevencmmib='177
 \font\fivecmmib=cmmib5 \skewchar\fivecmmib='177
 \newfam\cmmibfam
 \textfont\cmmibfam=\tencmmib \scriptfont\cmmibfam=\sevencmmib
 \scriptscriptfont\cmmibfam=\fivecmmib
 
\def\veps{\varepsilon}
\def\TT{{\mathcal T}}

\title{Algebras, Regions, and Observers}

 \author{Edward Witten}
\affiliation{School of Natural Sciences, Institute for Advanced Study,\\ 1 Einstein Drive, Princeton, NJ 08540 USA}
\abstract{In ordinary quantum field theory, one can define the algebra of observables in a given region in spacetime, but in the presence of gravity, it is expected that
this notion ceases to be well-defined.   A substitute that appears to make sense in the presence of gravity and that also is more operationally meaningful is to consider
the algebra of observables along the timelike worldline of an observer.   It is known that such an algebra can be defined in quantum field theory, and the timelike tube theorem
of quantum field theory suggests that such an algebra is a good substitute for what in the absence of gravity is the algebra of a region.   The static patch in de Sitter space is
a concrete example in which it is useful to think in these terms and to explicitly incorporate an observer in the description.
}

\begin{document}\maketitle

\section{Introduction}\label{intro} 

In ordinary quantum mechanics, we do not usually incorporate the observer as
part of the system.   
That is fine for many purposes, but in the presence of gravity, one has to take into account the fact
that the observer gravitates.     In some contexts, this is unimportant, because the observer's gravity  is
negligible.  For example, in an asymptotically flat spacetime, an observer more or less at rest in the asymptotic region is likely to have negligible gravitational
influence on whatever is in the interior of the spacetime.
   On the other hand, in  a closed universe, where the gravitational flux due to the observer has ``nowhere to go,'' one might expect
that it may be essential to include
the observer in the description.   

We will discuss
 an example of this -- based on the paper \cite{CLPW}.   But we begin with more general considerations.    In ordinary quantum field theory, one can attach an algebra $\A_\U$
of observables to a rather general  open set $\U$ in spacetime.   In the presence of gravity, there are potentially two problems with this notion.   First,
 when spacetime fluctuates, we may generically  have difficulty saying just what we mean by the spacetime region $\U$.   Second, it is not clear what is the logic of discussing an
algebra $\A_\U$ unless there is someone who can make the observations corresponding to elements of that algebra.   In the absence of gravity, we do not usually worry about this question; we just assume, in effect, that some observer external to the system has the relevant 
capability.    But in the presence of gravity, anticipating that sometimes it will be necessary to include the
observer in the description, we should take seriously the idea that it is only well-motivated to discuss an algebra of observables if this is the algebra of observables
accessible to someone.

But what algebra is accessible to an observer?  
We will discuss this question in the context of ordinary quantum field theory, without gravity, but hoping to learn some lessons that are useful when gravity
is included.   We  make use of two classic but not so well known results about ordinary quantum field theory.   The first result \cite{Borch}, described in 
section \ref{access},  enables one
to define an algebra of observables along the timelike worldline of an observer.   The second result, described in section \ref{ttube},
is the ``timelike tube theorem'' \cite{Borch2,Araki,Stroh,SW}, which among other things says that, in the absence of 
gravity, in a real analytic spacetime,
the algebra of observables along the observer's worldline is the same as (roughly) the algebra of observables in the region causally accessible to the observer.   The first
result is more elementary, and more general (since it does not require a hypothesis of real analyticity), but in a real analytic spacetime, it can potentially be viewed as a corollary
of the second.

The two results indicate that instead of discussing the algebra of observables in a spacetime region, we could discuss the algebra generated by the quantum fields
along the observer's worldline.
In a theory of gravity, the algebra of observables along the worldline would appear to be both better defined and operationally more meaningful than the algebra of a region.

In section \ref{desitter}, following \cite{CLPW}, 
we consider the ``static patch'' in de Sitter space as an example 
in which it is necessary to explicitly include an observer in order to define a sensible algebra
of observables.    The algebra defined after taking the observer explicitly into account turns out to be a von Neumann algebra of Type II$_1$.   This gives an abstract explanation
of why ``empty de Sitter space'' is a state of maximum entropy,   and in what sense the density matrix of empty de Sitter space is maximally mixed.   

Presumably, in a full theory of the world, an observer cannot be added from outside but must emerge as part of the theory.   In that context, what it means to include an observer in the
description is that one considers a ``code subspace'' of states in which the observer is present, and one defines operators that are well-defined on the code subspace, though they
would not be well-defined on all states of the theory.  For considerations of the present article, however, however, it is not necessary to have such a full theory in hand.

 \begin{figure}
 \begin{center}
   \includegraphics[width=2.5in]{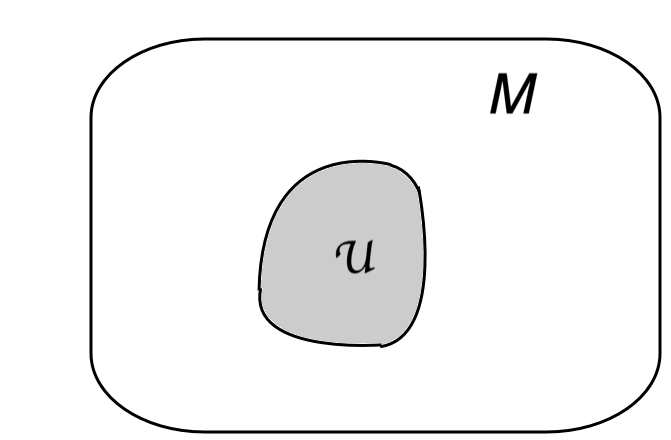}
 \end{center}
\caption{\footnotesize  In the absence of gravity, one can consider any chosen open set $\U$ in a spacetime $M$, and define the algebra of observables in that region. \label{one}}
\end{figure} 

\section{The Algebra Accessible to an Observer}\label{access}

 In ordinary quantum field theory in a spacetime $M$, we can arbitrarily specify any open set $\U\subset M$  and define an algebra
 $\A_\U$ of operators in $\U$ (fig. \ref{one}).  In the presence of gravity, since spacetime fluctuates, it does not make sense to talk
 about the region $\U$ unless we have an invariant way to identify it.   For example, in an asymptotically flat spacetime, in the presence of a black hole, we could talk about the region outside the black hole
horizon.      That is invariantly defined and presumably makes sense at least perturbatively even when spacetime fluctuates. 
We could introduce an observer who is more or less at rest near infinity and describe the region outside the horizon as the region visible to this observer.
 However, as  already noted, in an asymptotically flat  universe we do not expect that it is essential to incorporate the observer in the description.  
 
\begin{figure}
 \begin{center}
   \includegraphics[width=2.5in]{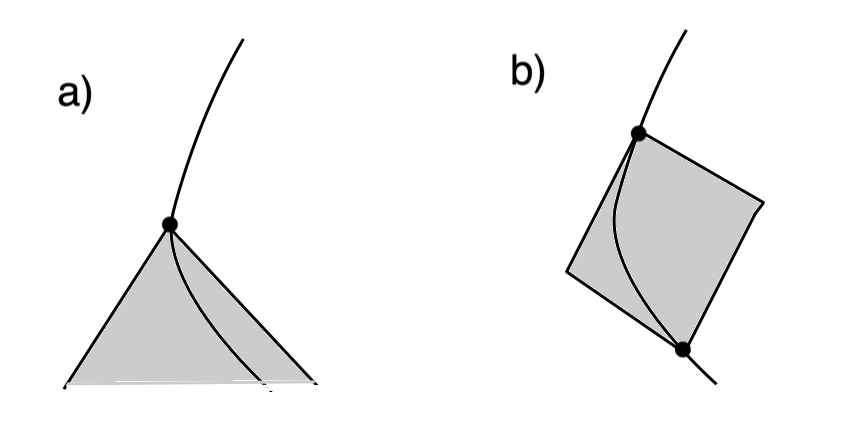}
 \end{center}
\caption{\footnotesize  The worldline of an observer, with (a) the region of spacetime that is visible to the observer prior to a given time, or (b) the region that the observer
can both see and influence in a given time interval.  Time runs vertically and the boundaries of the chosen time intervals are marked by the black dots. \label{two}}
\end{figure}

 If we assume the existence of an observer, we can invariantly identify various regions in spacetime.  For example, if the observer
carries a clock, then as in fig. \ref{two}, we can discuss the region that is visible to the observer prior to a given time, or, alternatively, 
the region that is causally accessible to the observer in a stated time interval  (meaning that the observer
can both see and influence this region during the interval in question).  

But what can an observer actually measure?    Here we will assume a very simple model in which the observer is described by a timelike worldline, and what the observer
can measure are simply the quantum fields along this worldline.\footnote{See \cite{Unruh} for a classic discussion in this framework, with a somewhat different motivation.}      
  This seems like a rather minimal model of what an observer is, and one could well worry that it is too crude.  
   A realistic observer would presumably also carry measuring equipment, and a recording device,
and would have access to operators that act on all that.
  But it turns out that in ordinary quantum field theory
  without gravity, the rather crude model in which an observer is just characterized by a worldline  and the observables are the quantum fields along the worldline
  is  sufficient, for many purposes.
  
This model raises two immediate questions:

(1)  Can well-defined operators be defined by smearing a quantum field along a timelike worldline?

(2)   Given a ``yes'' answer to the first question, what  is the algebra generated by these operators? 

Let us elaborate a bit on the first question. We are accustomed  in quantum field theory  to talking
about ``local operators'' $\phi(x)$, but a local operator is not really a Hilbert space operator,
since acting on a Hilbert space state it takes us out of Hilbert space. 
In the case of the vacuum state $\Omega$ in Minkowski space, this is
 clear from the fact that $|\phi(x)|\Omega\r|^2=\infty$ or equivalently
\be\label{firstone} \l\Omega|\phi(x) \phi(x)|\Omega\r=\infty, \ee
due to a short distance singularity.     Since the leading short distance singularity is universal, 
it is also true that $|\phi(x)|\Psi\r|^2=\infty$ for any state $\Psi$ in any spacetime $M$.  The problem has nothing to do with the state
$\Psi$, and purely reflects the fact the product $\phi(x)\cdot \phi(x)$ is not well-defined, since in fact a more general product
$\phi(y)\cdot \phi(x)$ is singular for $y\to x$.   This singularity is governed by the operator product expansion (OPE).

If we could measure $\phi(x)$, the answer would be one of its eigenvalues, but since $\phi(x)$ maps us out of Hilbert space, it does not have eigenvectors or eigenvalues,
and we cannot measure it.   
What are actually measureable are suitable smeared versions of $\phi(x)$.    Which ones?
Suppose we are going to smear a real scalar field $\phi(x)$ over a set $S$ to get a smeared ``operator''
$\phi_f=\int_S \d \mu f(x)  \phi(x),$ where $f(x)$ is a complex-valued smooth function with support in $S$.
If $\phi_f$ is actually going to make sense as an operator, the smearing has to be such that the $\phi(x')\cdot\phi(x)$
OPE singularity is integrable, when smeared in this fashion.
 
 For example, spatial smearing will only succeed for an operator of rather low dimension.   In $\sD=d+1$ spacetime dimensions, with space coordinates $\vec x$ and a time coordinate $t$,
 spatial smearing at, say, $t=0$, produces an expression
 $\phi_f=\int_{t=0}\d^d \vec x f(\vec x)\phi(\vec x,0)$, where $f(\vec x)$ is a smooth  function of the spatial coordinates $\vec x$ that we can assume to have compact support.  Does
 $\phi_f$ make sense as an operator?
In computing the product $\phi_{\bar f}\cdot \phi_f$, we run into the integral
 \be\label{opprod} \int_{t'=t=0}\d \vec  x\d\vec x' \bar f(\vec x') f(\vec x) \phi(\vec x',0) \phi(\vec x,0). \ee
For illustrative purposes, let us consider first the case of a conformal field theory, though  with minor modifications, the following remarks apply much more widely.
   If $\phi$ is a conformal field of  dimension $\Delta$, then 
 the  leading singularity in the operator product $\phi(\vec x',0) \phi(\vec x,0)$
 for $\vec  x'\to \vec x$ is proportional to $|\vec x'-\vec x|^{-2\Delta}$, where $\Delta$ is the dimension of the operator $\phi$.   
  The condition for this singularity to be integrable when inserted in (\ref{opprod}) is 
 $2\Delta<d$.    If $\phi$ is a free scalar field, then $\Delta=(\sD-2)/2=(d-1)/2$, and the condition $2\Delta<d$ is satisfied.\footnote{In dimension $\sD=2$, a free scalar has $\Delta=0$
 and the condition $\Delta<d/2$ is satisfied by any normal ordered polynomial $:\phi^n(x):$.   This was important in early work on constructive field theory \cite{Jaffe}.} 
 But what about, say, QCD, in the real
 world with $d=3$?    QCD is asymptotically free, so short distance singularities have the behavior just discussed up to logarithms, which are inessential except in the borderline
 case $\Delta=d/2$.  In QCD,
 the smallest value of $\Delta$ for any gauge-invariant operator is 3, corresponding to a quark bilinear such as $\bar q q$.  So the condition $2\Delta<d=3$
 is never satisfied and QCD is an example of
 a theory in which no true operator can be produced by smearing of a ``local operator'' in space.
 
 Smearing in Euclidean space is only slightly better.  If we try to define a smeared operator
 $\phi_f=\int\d^\sD x f(x) \phi(x)$ (where now the integration is over all $\sD$ coordinates of Euclidean spacetime),  
 we will run into the operator product singularity $\int\d^\sD x \d^\sD x' \bar f(x')f(x) \phi(x')\phi(x).$
 This is integrable if and only if $2\Delta < \sD$, a weaker condition but one that again cannot be satisfied in QCD, for example.
 
 How then do we get true operators by smearing of ``local operators''?   The secret is smearing in real time. Though smearing in space is only effective in favorable cases,
  smearing in real time turns
 a ``local operator'' of any dimension into a true operator.   This rather old result \cite{Borch} was originally proved directly on the basis of the Wightman axioms of quantum
 field theory for the case that the timelike curve is a timelike
 geodesic in Minkowski space.     
 
 To understand the result, we begin again with the case of a conformal field theory.
 Suppose that at $\vec x=0$, we smear a ``local operator'' $\phi(\vec x,t)$ by a compactly supported function $f(t)$ that depends only on $t$.   
Thus we define $\phi_f(0)=\int \d t \,f(t) \phi(0,t)$.   In evaluating $|\phi_f|\Omega\r|^2$, we now run into the operator product
$\int \d t'\d t \bar f(t') f(t)\phi(0,t')\phi(0,t)$. 
If $\phi$ has dimension $\Delta$,  the leading
OPE singularity is $\phi(t')\phi(t)\sim (t'-t-\i \veps)^{-2\Delta}$, so we have to consider the integral
\be\label{goodint} \int \d t'\, \d t \,\bar f(t') f(t) \frac{1}{(t'-t-\i\veps)^{2\Delta}}.\ee
There are also subleading OPE singularities; they have the same form with different exponents and can be treated just as we are
about to describe.      The integral is obviously well-defined  for $\veps>0$, and we want to show that no divergence appears in the limit
$\veps\to 0$.   For this,  we write\footnote{If $2\Delta\in\Z$, the following formula has to be slightly modified with a logarithmic factor on the
right hand side.   The derivation otherwise proceeds in the same way.}
\be\label{justwr} \frac{1}{(t'-t-\i\veps)^{2\Delta}}=C_n \frac{\partial^n}{\partial t^n} (t'-t-\i\veps)^{n-2\Delta},\ee
for any integer $n>0$, with a constant $C_n$.    Inserting this in the integral (\ref{goodint})  and integrating by parts $n$
times, we replace the original integral with
\be\label{noint} (-1)^n C_n \int \d t' \,\d t \,\bar f(t') f^{[n]}(t)  \, (t'-t-\i\veps)^{n-2\Delta}.\ee
For large enough $n$, this is manifestly convergent for $\veps\to 0$.

In a general quantum field theory, consider the operator product expansion 
 \be\label{ope}\phi(0,t')\phi(0,t)\sim\sum_\alpha h_\alpha(t'-t-\i\veps)\phi_\alpha(0,t).\ee  The coefficient functions $h_{\alpha}(t'-t-\i \veps)$ are holomorphic
for  $\Im(t'-t)<0$. This follows from positivity of energy.  Normally one can assume that the 
singularities of the $h_\alpha$ are bounded\footnote{An example of a local
operator that would not satisfy this condition is $e^{\i \phi(x)}$, where $\phi$ is a scalar field in spacetime dimension $\sD>2$.}   by a power law,  $|h_\alpha(t'-t)|<C_\alpha|t'-t|^{-n_\alpha}$, for
some $C_\alpha,n_\alpha$.   In the context of the Wightman axioms of
quantum field theory,  the correlation functions are usually assumed to be tempered distributions, which implies such a bound.  Alternatively, if a theory is conformally invariant or asymptotically free in the ultraviolet, and the operators considered behave as fields of definite dimension in the ultraviolet
(modulo logarithms in the asymptotically free case), this again implies such a bound.   Given holomorphy 
in the lower half $t'$ plane and a power law bound,\footnote{The precise mathematical statement is 
that if $f(z)$ is a function holomorphic in the lower half
plane, then a necessary and sufficient condition for the boundary values of $f(z)$ along the real axis to define a distribution is a bound $|f(z)|<k|\Im\,z|^{-c}$ for some constants $k,c$. 
For an elementary proof (and a bound on the distribution) see Proposition 4.2 in \cite{BF}.
See also Theorem 1.1 in \cite{Straube} for a $d$-dimensional generalization.}   
 one can imitate the previous derivation to show the finiteness of 
\be\label{zelbo}\int \d t'\d t \bar f(t') f(t)\phi(0,t')\phi(0,t)\sim \lim_{\veps\to 0}\sum_\alpha \int \d t'\d t \bar f(t')f(t) h_\alpha(t'-t+\i\veps)\phi_\alpha(t).\ee For this, one writes each $h_\alpha(t'-t-\i\veps) $ as 
the $n^{th}$ derivative with respect to $t'$, for some $n$, of a function $k_\alpha(t'-t-\i\veps)$
that remains continuous (though not smooth) for $\veps\to 0$, and then one integrates by parts as in the derivation of eqn. (\ref{noint}).

What happens if  we consider an arbitrary timelike curve $\gamma$ in Minkowski space, not necessarily a geodesic? Parametrize $\gamma$
 by the proper time $\tau$ and let $d(\tau',\tau)$ be the signed
proper distance\footnote{The proper distance  in Minkowski space between points along $\gamma$ labeled by $\tau'$ and by $\tau$ 
 is the proper time elapsed along a geodesic between the two points (not along the path $\gamma$).  To get the signed
proper distance $d(\tau',\tau)$, we multiply the proper distance by $+1$ if $\tau'>\tau$ and by $-1$ if $\tau>\tau'$.} in Minkowski space between points on 
$\gamma$ labeled by $\tau'$ and by $\tau$.   The effect of replacing a timelike geodesic by an arbitrary timelike curve is to replace $\tau'-\tau-\i\veps$ in the preceding formulas by 
$d(\tau',\tau)-\i\veps$.   Since $d(\tau',\tau)$ is a smooth function and $d(\tau',\tau)\sim \tau'-\tau+\O((\tau'-\tau)^2),$ 
this does not substantially affect the preceding analysis.  The singularities at $\tau'\to \tau$ are of the same general form and are harmless.

What happens if we replace Minkowski space by a general spacetime $M$?    Intuitively, one would not expect this to matter, since everything is determined by   short distance
behavior.   In a curved spacetime, the operator product expansion becomes more complicated, with curvature dependent terms; see \cite{HW} for an axiomatic discussion.   But the 
extra terms have similar singularities to what we have already considered, so one would expect the same result.  For a proof,
under  reasonable assumptions about  correlation functions in a curved spacetime, see \cite{K}.  See also further discussion in \cite{Fewster},  \cite{SW}.

\section{The Timelike Tube Theorem}\label{ttube}

\subsection{The Timelike Envelope and the Timelike Tube Theorem}\label{envelope}

In section \ref{access}, we learned that one can define operators by smearing a quantum field along the timelike worldline of an observer.   Therefore  we can consider
the algebra generated by such operators (or more precisely by bounded functions of such operators).  
But what are the algebras that we make this way?   In the context of quantum field theory without gravity, this question
is answered by the ``timelike tube theorem.''  This theorem was originally formulated for timelike geodesics in Minkowski space \cite{Borch2,Araki}.
It was generalized to free field theories in curved spacetime in \cite{Stroh}. For a version of the theorem suitable for non-free theories in
curved spacetime, see \cite{SW,SW2}.

 \begin{figure}
 \begin{center}
   \includegraphics[width=2.5in]{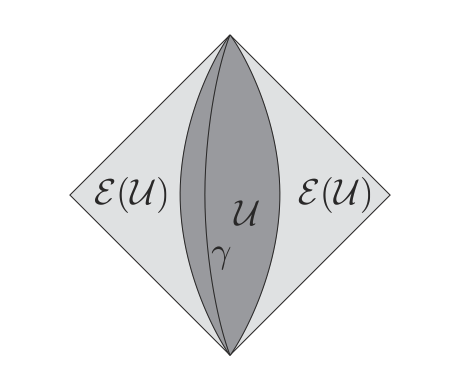}
 \end{center}
 \caption{\footnotesize The timelike envelope $\E(\U)$ of an open set $\U$ consists of all points that can be reached by  deforming a timelike curve in $\U$ keeping its
 endpoints fixed. In general, one considers all possible timelike curves in $\U$, but in the case depicted, it suffices to consider  segments of  the particular
 timelike curve  $\gamma$.   \label{three}}
\end{figure}  

If $\U$ is an open set in spacetime, its ``timelike envelope'' $\E(\U)$ consists of 
all points that can be reached by deforming timelike curves in $\U$ through a family of timelike curves, keeping the endpoints
fixed (fig. \ref{three}).  (Thus $\E(\U)$ is contained in the intersection $J^+(\U)\cap J^-(\U)$ of the past and future of $\U$, but in general it is smaller.
See section \ref{stronger} for an example.)   The timelike tube theorem asserts that the algebra of operators\footnote{For a more precise explanation of what algebra is intended here, see section \ref{additive}.}  in $\U$ is the same as the algebra of operators in the possibly much
larger region $\E(\U)$.    We will aim to give at least a hint of why this is true, after first explaining
an implication, noted in \cite{Stroh}.   

Suppose that we are actually interested in a timelike curve $\gamma$, possibly  one of finite extent
with  endpoints $q,p$ or possibly an infinite or semi-infinite curve.   
The timelike envelope $\E(\gamma)$ consists of all points that can be reached by deforming $\gamma$ through a family of timelike curves,
keeping $\gamma$ fixed near its ends.   If $\gamma$ has endpoints, they are omitted from $\E(\gamma)$ to ensure that $\E(\gamma)$ is an open set.
 We can thicken $\gamma$ (minus its endpoints, if any)
  to an open set $\U$, in such a way that the timelike envelope $\E(\U)$ does not depend on $\U$ -- it is the same as the timelike envelope $\E(\gamma)$
 of $\gamma$. See fig. \ref{three}.
The timelike tube theorem says that the algebra $\A(\U)$ of operators in region $\U$ does not really depend on $\U$ but only on $\gamma$.   It coincides with $\A(\E(\gamma))$, the
algebra of operators in $\E(\gamma)$.   So we can define an algebra for every (possibly bounded) timelike curve $\gamma$.  
 That in itself is not a surprise, since we arrived at the same conclusion in a more direct way\footnote{The timelike tube theorem depends on real analyticity of spacetime,
 as discussed shortly, and the reasoning in section \ref{access} does not, so the previous analysis was more general.} in section \ref{access}.   But it is nice to see the consistency with what we learned from
 more elementary arguments.
 
 Strictly speaking, the algebra associated to a curve $\gamma$ in section \ref{access} consists of the quantum fields smeared along $\gamma$, while
 the timelike tube theorem gives us an algebra $\A(\gamma)$ of operators that can be defined in an arbitrarily small neighborhood of $\gamma$.   Presumably, these two algebras coincide
 (assuming that we include all possible local operators in the construction of section \ref{access}, including derivatives of operators).
 A proof might start with an axiomatic characterization of local operators at a point in terms of a limit of the operators in a small ball around that point.
  Were the two algebras to differ, one might argue that as the notion of an observer characterized by an infinitely thin worldline is
 an idealization, the physically more relevant algebra would be the one provided by the timelike tube theorem.

The interpretation that we want to give is that the algebra $\A(\gamma)$ of operators
supported on a curve $\gamma$ is a good stand-in for the algebra $\A(\E)$ of the spacetime region $\E$ associated to $\gamma$.  The algebra $\A(\gamma)$ is more operationally
meaningful than $\A(\E)$, since it is more directly what an observer can measure.   And it is also better defined in the presence of gravity than $\A(\E)$ would appear to be.
Once one takes into account fluctuations in spacetime, it is hard to see how one would directly define $\A(\E)$, but as long as an observer is included in the description,
the algebra of observables along the observer's worldline seems to be a meaningful notion, even when the observer's worldline fluctuates.   So $\A(\gamma)$ 
seems like a good substitute for the algebras associated to open sets
that one  considers in the absence of gravity.

\begin{figure}
 \begin{center}
   \includegraphics[width=3.5in]{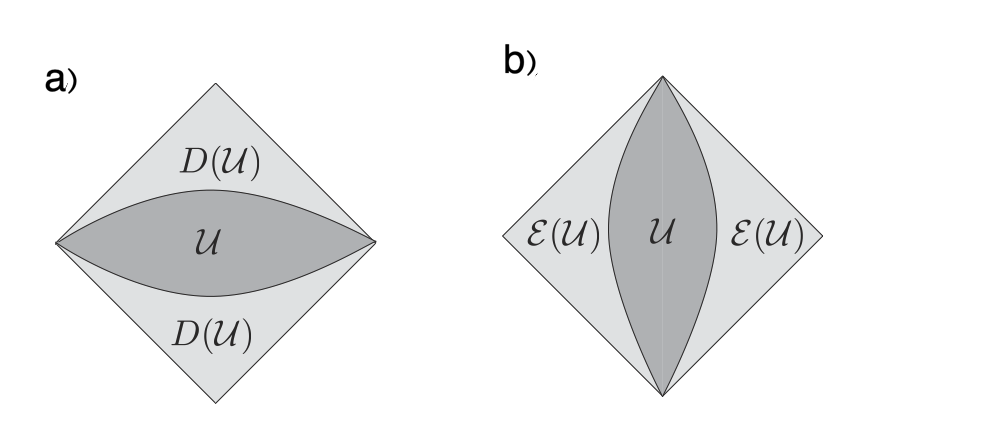}
 \end{center}
 \caption{\footnotesize Two cases in which the solution of a hyperbolic wave equation is given in one open set  (the dark shaded region) 
 and one wishes to extend the solution over a larger
 open set (the more lightly shaded region). Time runs vertically and space runs horizontally.  In (a), the solution is given in a ``spacelike pancake'' $\U$ and one wishes to extend it over the domain
 of dependence $D(\U)$.   In (b), the solution is given in a ``timelike tube'' $\U$, and
 one wishes to extend it over the ``timelike envelope'' $\E(\U)$.  In dimension 2, there is a perfect symmetry between these two cases, but
in higher dimension, there is no such symmetry.    \label{four}}
\end{figure}

To try to gain some idea of why the timelike tube theorem is true, let us consider the classical limit.    Suppose
we have a reasonable relativistic wave equation like the Klein-Gordon equation $(\Box+m^2)\phi=0$, where $\Box$ is the wave operator, or possibly a nonlinear modification of this equation. 
In fig. \ref{four}, we are given a solution in one region of a spacetime $M$,
and we want to predict the solution in a larger region.   We consider two cases.  In fig. \ref{four}(a),  the solution is given in a ``spacelike pancake'' $\U$, and we want to
extend it over the domain of dependence $D(\U)$ of  $\U$.     In fig. \ref{four}(b), the solution is given in a ``timelike tube'' $\U$  and we want to extend 
the solution over the corresponding ``timelike envelope'' $\E(\U)$.

\begin{figure}
 \begin{center}
   \includegraphics[width=2.5in]{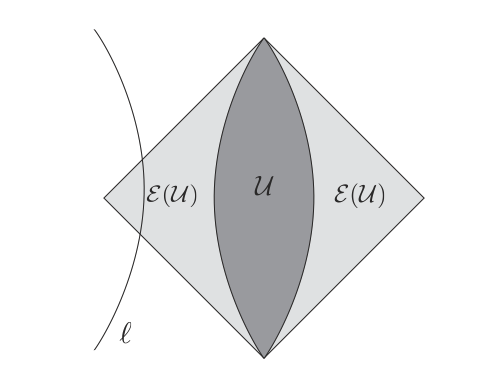}
 \end{center}
 \caption{\footnotesize This picture illustrates the obstruction to existence in the setting of fig. \ref{four}(b).    Starting with initial data in an open set $\U$, if one tries to extend
 a given solution in spatial directions over the timelike envelope $\E(\U)$, it may develop singularities that will prevent the existence of the extended solution.  In fact,
 this is the generic behavior.  The singularity might arise on the worldline of a point charge that passes through $\E(\U)$ but not through $\U$, as sketched here. \label{five}}
\end{figure}

 There is a basic  asymmetry between the two cases.\footnote{\label{special} To be precise, this asymmetry holds in any 
 spacetime dimension $\sD>2$.  In $\sD=2$, spacetime has one dimension of space
 and one dimension of time, and there is a perfect symmetry between the two cases of fig. \ref{four}. In $\sD>2$, there are more space dimensions than time dimensions,  and there is no
 such symmetry.     The counterexample illustrated in fig. \ref{five} fails in  $\sD=2$ because in $\sD=2$, the solution with a delta function source along a worldline $\ell$ does
 not blow up along $\ell$, but instead is discontinuous across $\ell$.  Starting with the solution on one side of $\ell$,
 one can drop the discontinuity and smoothly continue the solution across $\ell$.  In $\sD>2$, a solution with such a delta function source blows up
 along $\ell$ and there is no way to remove the singularity without
 changing the solution in the original region $\U$.}   The Holmgren uniqueness theorem of classical partial differential equations\footnote{For an accessible exposition of this theorem,
 see chapter 5 of \cite{Smoller}.}  asserts that the extension over the larger region is
 unique, if it exists, both in fig. \ref{four}(a) and in fig. \ref{four}(b).    But existence is more special and only holds in fig. \ref{four}(a).

A simple counterexample (fig. \ref{five}) shows that an existence result cannot possibly hold in fig. \ref{four}(b).
Let $\ell$ be a timelike curve that passes through the region $\E(\U)$ but not through $\U$.  Consider a solution of the equation $(\Box+m^2)\phi=0$ with a delta function source
along $\ell$.   If we consider Maxwell's equations rather than the Klein-Gordon equation, then $\ell$ can be the worldline of a point charge.    The field 
$\phi$ blows up along $\ell$ (in dimension $\sD>2$; see footnote \ref{special}).  Starting with the solution in the region $\U$, there is no way to
extend it over $\E(\U)$ as a solution, since if one tries to do this, one encounters the blowup along $\ell$.   The equation is not obeyed along $\ell$, because of the delta
function source.  

The existence and uniqueness result in the case of fig. \ref{four}(a) is the basis for much of physics.   It says that the solution can be predicted from initial data --
physics is causal.   But by contrast the uniqueness result without
a guarantee of existence in fig. \ref{four}(b)   is not usually useful at the classical level because generically the extension over $\E(\U)$ of a solution on $\U$  does not exist and it
is very hard to predict when it does.

Suppose, however, that  we are doing quantum field theory and for simplicity consider a free field $\phi$ with the action\be\label{zoldo} I=\frac{1}{2}\int_M \d^\sD x \sqrt g\left(-D_\mu\phi D^\mu\phi -m^2\phi^2\right).  \ee
  In this case, we can view $\phi$ as an operator-valued solution of the Klein-Gordon equation $(\Box+m^2)\phi=0$.  
If we are studying this quantum field theory on $M$, then the field $\phi(x)$ does exist throughout $M$ and therefore existence of
the extension from $\U$ to $\E(\U)$ is not an issue.

But what does uniqueness mean?   In some sense, uniqueness means ``the field $\phi(x)$ for $x\in \E(\U)$ is uniquely determined by $\phi(y)$ for $y\in\U$.''
 As explained by Borchers and Araki in the early 1960's, the quantum meaning of this statement is really ``$\phi(x)$ for $x\in \E(\U)$
is contained in the algebra generated by $\phi(y)$ for $y\in \U$'' or equivalently the operator algebras of the two regions are the same:
$$\A(\E(\U))=\A(\U).$$
This is the timelike tube theorem.

Classically, we could just as well add higher order terms to the action and consider a field that satisfies a nonlinear partial differential equation.  Holmgren uniqueness applies equally well 
to such an equation.
Quantum mechanically, although a free quantum field can be viewed as an operator-valued solution of a classical field equation, that is not the case for a 
non-free quantum field, because of issues involving renormalization.    Accordingly, proofs of the timelike tube theorem in non-free theories require  additional ingredients,
beyond what is needed in free theories.  An approach that works for non-free theories in curved spacetime was presented in  \cite{SW} (see \cite{SW2} for an informal account).

In this discussion, we have skipped over two important points.   First, we have not been precise about what is the algebra  that is governed by the timelike tube theorem.
See  section \ref{additive}.    Second, we have omitted to explain a key assumption in the classical Holmgren uniqueness theorem and the quantum timelike tube
theorem.  These theorems hold in real analytic spacetimes.   Real analyticity is not needed for existence and uniqueness in the setting of fig. \ref{four}(a), but it is needed for
uniqueness in the setting of\footnote{Actually, it suffices for the spacetime to be, in some coordinate system, real analytic in the time direction \cite{Tataru}. In free field theory,
this is also true for the timelike tube theorem \cite{Stroh}.}  fig. \ref{four}(b).
Is this restriction important for the application of the timelike tube theorem to semiclassical gravity?   In semiclassical gravity, one is usually studying the behavior of quantum 
fields in a background spacetime that in practice is normally real analytic.   For instance, in section \ref{desitter}, we consider the static patch in de Sitter space as an example in 
which it is important to explicitly include an observer in order to define a sensible algebra of observables, and thus in which the timelike tube theorem provides important motivation.
This is a typical example in which the starting point is a real analytic spacetime.  After picking a semiclassical starting point, one then expands around it.   
In perturbation theory, the metric certainly fluctuates away from the real analytic starting point, but one expects that 
in perturbation theory, the gravitational field can be treated like any other quantum field in a real analytic background.   
Thus it appears that for typical applications of the timelike tube theorem to semiclassical gravity, the restriction of the theorem
to real analytic spacetimes is not a problem.

\subsection{The Additive Algebra}\label{additive}  

For a region $\U$ in spacetime, what precisely is the algebra of operators to which the timelike tube theorem applies?
The smallest reasonable candidate is the algebra generated by local operators in $\U$ (or more precisely, by bounded functions of smeared local operators).   
 If  $\U$ is contractible, this is the natural algebra of observables in
$\U$ in a generic quantum field theory.     However, if $\U$ is not contractible, then in general, as we discuss shortly, there can be
 additional operators in $\U$, such as Wilson operators defined on noncontractible loops in $\U$, that cannot be constructed from local operators in $\U$.  
 
 The algebra generated by local operators has been called the additive algebra $\A_\add(\U)$ \cite{casini1,casini2}.   One can then reserve the name $\A(\U)$ for the possibly
 larger algebra of all operators in $\U$. The timelike tube theorem is really a theorem about $\A_\add(\U)$.   Thus a more precise statement of the theorem than we have given
 so far is $\A_\add(\U)=\A_\add(\E(\U))$.  This is clear from the proofs, which involve studying algebras of local operators. 
  However, as we will discuss, while the distinction between $\A(\U)$ and $\A_\add(\U)$ exists in quantum field theory in general,
 there are reasons to believe that this distinction is absent in any  theory that emerges at long distances from  a full model of quantum gravity.
 
 The motivation for the phrase ``additive algebra'' is as follows.
 Let $\B$ be a small open ball in $\U$.  Then as $\B$ is contractible, there is no distinction between $\A(\B)$ and $\A_\add(\B)$.    Now cover $\U$ by open balls $\B_\alpha$, with
 $\alpha$ ranging over some set $S$.   The additive algebra of $\U$ is the same as the algebra generated by the $\A(\B_\alpha)$, $\alpha\in S$:
 \be\label{melvin} \A_\add(\U)=\vee_{\alpha\in S}\A(\B_\alpha). \ee

 In what situation will we have $\A(\U)\not=\A_\add(\U)$, where $\A(\U)$ is the algebra of all possible operators in $\U$, not necessarily built from local operators?  
 For a typical example, consider a theory with a $\UU(1)$ gauge field $A$, with curvature $F=\d A$.   Let  $\ell$ be a closed
 loop in $\U$, and consider the Wilson operator $W_\ell(A)=\exp(\i\oint_\ell A)$.  Is this operator contained in $\A_\add(\U)$?    The answer will be ``yes'' if $\ell$ is the boundary
 of an oriented two-manifold $D\subset \U$, for then $W_\ell(A)=\exp(\i\int_D F)$ is a bounded function of a smeared local field in $\U$, namely $\int_D F$.    But even if $\ell$ is not
 a boundary in $\U$, the answer can still be ``yes,'' for a more subtle reason that was explained in \cite{Harlow}. Consider a theory that has a charge 1 field $\phi$.   Then for an open path
 $\gamma\subset \U$, say with endpoints $q,p$, one can consider the gauge-invariant operator
 $V_\gamma(A)=\bar\phi(p) \exp(\i\int_\gamma A)\phi(q).$   In the case that $\gamma$ is a very short path, with $p$ very near $q$, $V_\gamma$ can be expanded in terms of local
 operators at $q$ and so is contained in the additive algebra of any open set containing $\gamma$.    If $\ell$ is a closed loop, then $W_\ell(A) $ can be ``cut'' in the
 sense that if one omits from $\ell$
  a number of open intervals of size $\veps$, with the portion retained then being a disjoint union of closed intervals $\gamma_i$, $i=1,\cdots ,n$, then $W_\ell(A)$ appears
 as the most singular contribution
 in the operator product $\prod_{i=1}^n V_{\gamma_i}(A)$ for $\veps\to 0$.    Hence, in this situation $W_\ell(A)$ is contained in the additive algebra $\A_\add(\U)$ for any region $\U$ containing $\ell$.
 
 However, if the theory has no field of unit charge, and $\ell$ is not a boundary in $\U$, then $W_\ell(A)$ is contained in $\A(\U)$ but not in $\A_\add(\U)$.   This phenomenon
 is certainly not limited to $\UU(1)$ gauge theory.
In a theory with any gauge group $G$, if some representation $R$ of $G$ is ``missing,'' in the sense that no local operator of the theory transforms
 in this representation,  then for suitable regions $\U$ one will have  $\A_\add(\U)\not=\A(\U)$. A Wilson loop in the representation $R$ can be present in $\A(\U)$ but not in $\A_\add(\U)$.
   A similar role can be played by a $p$-form gauge field $B$ with $p>1$. In a theory that
 has such a field, one can consider the operator $Q_S=\exp(\i\int_S B)$, where $S$ is a $p$-cycle in $\U$.   If $S$ is not a boundary in $\U$ and the theory does not have a 
 string or membrane  that couples to $B$ (which would enable one to ``cut'' the operator $Q_S$ in small pieces), then $Q_S$ is contained in $\A(U)$  but not in $\A_\add(\U)$. 
 There is also an analog of this for magnetic charges (and their analogs for strings and membranes), with 't Hooft operators instead of Wilson operators.    
 
 We can now easily give an example that shows that it is necessary to specify that the timelike tube theorem applies to $\A_\add$, not to $\A$.   Consider $\UU(1)$ gauge theory
 without charged fields on a spacetime $\R\times S^1$ with product metric, where $\R$ parametrizes time and $S^1$ is parametrized by an angular variable $\phi$.  In this spacetime,
 there is a nontrivial Wilson operator $W_\ell=\exp(\i \oint_\ell A)$, where $\ell$ is a closed loop that wraps around the $S^1$.   Let $p$ and $q$ be two points
 at the same value of $\phi$ but different values of $t$, and let $\gamma$ be the timelike geodesic that runs from $q$ to $p$ at fixed $t$.  
If $\U$ is a small neighborhood of $\gamma$, there is no nontrivial Wilson operator in $\U$, so $\A(\U)=\A_\add(\U)$.   But if $p$ is  separated in time from $q$ by more than the circumference of the $S^1$, then the region $\E(\U)$ includes a circle that wraps all the way around the $S^1$, so $\A(\E(\U))\not=\A_\add(\E(\U))$.   Thus in this example, we
have $\A_\add(\U)=\A_\add(\E(\U))$, as guaranteed by the timelike tube theorem, but $\A(\U)\not=\A(\E(\U))$.
  
Let us say that a quantum field theory which may have  gauge fields or $p$-form gauge fields for $p>1$ is ``complete'' if the electrically and magnetically charged particles, strings, and membranes  coupling to those gauge fields are a maximal possible set consistent with all principles of low
energy physics including Dirac quantization of electric and magnetic charge.  
Complete theories are precisely the ones with $\A_\add(\U)=\A(\U)$ for all $\U$, so we might also call them ``additive.'' 
It is believed that a quantum field
theory that emerges at low energies from a full-fledged (ultraviolet complete) theory of quantum gravity is always complete in this sense.     This was originally suggested based 
 on experience with string theory \cite{Polchinski}.  Further arguments were given in \cite{BS}, and the fullest known argument is based on holographic duality \cite{HO}.  
 
The claim that a theory that emerges from a full quantum gravity theory will be additive or complete has another interpretation.
A theory with ``missing'' charges (or strings or membranes)   has
 a $p$-form global symmetry  in the language of \cite{GKSW} (where $p$ depends on what is missing).  So the statement that a full-fledged theory of quantum gravity
is complete can be viewed as an extension to $p$-form global symmetry with $p>1$ of the statement that there are no global symmetries
 in a full theory of quantum gravity.   For discussion of this statement, see for example \cite{BS,EW}.
 
 In the spirit of the present article, one can motivate as follows the idea that an ultraviolet-complete theory of quantum gravity should have no missing charges.
We assume that a complete gravity theory can be approximated, in an appropriate class of states, by  an ordinary quantum field theory weakly coupled to gravity.   Suppose that this ordinary quantum field theory is not complete, for example
because it has a $\UU(1)$ gauge field but no state of charge 1.   Then if a loop $\ell$ is not a boundary in spacetime, the Wilson operator $W_\ell=\exp(\i\oint_\ell A)$ is not
measurable by any observer living in the spacetime and described by the theory.   After all, given that $\ell$ is not a boundary, $W_\ell$ cannot be measured by measuring the curvature
$F=\d A$.  To measure $W_\ell$ when $\ell$ is not a boundary  requires measuring the interference
between different histories of a charge 1 particle that differ by the homology class of the worldline of the particle.   But the assumption that the theory has no state of charge 1 means
that such an experiment is not possible using the resources available in the theory.

What, then, do we mean in claiming that $W_\ell$ (or its hermitian part) is an ``observable''?    In ordinary quantum mechanics, we consider the observer to be external to the system,
and we make no particular assumption about what resources the observer can use in probing the system.  From that point of view, the observer might be able to introduce a massive
charged particle to probe the system and measure $W_\ell$.   The observer, after all, is described by a more complete theory of the universe that might have the necessary charged
particle.    An ultraviolet-complete theory of quantum gravity, however, is expected to describe all that there is in the universe that it describes, 
with no way  to add anything from outside.  In the context of such a theory, the observer and any apparatus
used by the observer must already be described by the theory.  So in an ultraviolet complete theory of quantum gravity with a ``missing'' charge, we would be in the
awkward situation that the theory would enable us to define an ``observable''  $W_\ell$ that would be well-defined in an appropriate, long distance limit, but that no 
one, in principle, could measure.
 
  \subsection{What Else?}\label{stronger}
  
  In addition to the timelike tube theorem, which asserts that $\A_\add(\U)=\A_\add(\E(\U))$, one has causality, which, denoting the domain of dependence of a set $\U$ as $D(\U)$,
  asserts that $\A(\U)=\A(D(\U))$.   If the difference between $\A$ and $\A_\add$ is unimportant, these statements can be combined, relating $\A(\U)$ to the algebra of a
  set that in general is  larger than either $\E(\U)$ or $D(\U)$.   This was noted by Araki \cite{Araki} in one of the original papers on the timelike
  tube theorem.   However, Araki actually conjectured a further generalization.   In the context of quantum fields in Minkowski space, Araki wrote, 
  `` ... If $\ell$ is a snake-like line, connecting two mutually timelike points $P_1$ and 
  $P_2$ but everywhere spacelike, and $\B$ is a tube around $\ell$, of a small diameter, then it is rather likely that any solution of the [massless scalar] wave equation vanishing in $\B$  
  might always vanish in the double light cone spanned by $P_1$, and $P_2$. If such a conjecture turns out to be true, then we immediately have the corresponding 
  theorem for $R(\B)$.'' In our terminology, the proposal is that if $\Delta$ is the double cone or causal diamond
  with vertices $P_1$ and $P_2$, then $\A_\add(\B)=\A_\add(\B\cup \Delta)$.     This follows from arguments in Araki's paper if it is true that any solution of the massless scalar wave equation
  vanishing in $\B$ also vanishes in $\Delta$.  This statement does not follow from Holmgren uniqueness in any 
  obvious way and its validity remains unclear sixty years later.
  
  In general, what
   extension of the timelike tube theorem is conceivable?  For any set $\U$ in a spacetime $M$, one writes $\U'$ for the set of points in $M$  that are spacelike separated from $\U$.
  Then $\U''=(\U')'$ is called the causal completion of $\U$.   For any open set $\U$, since 
  $\A_\add(\U)$ commutes with $\A_\add(\U')$, clearly any statement that $\A_\add(\U)=\A_\add(\V)$ for some open set $\V$ implies that  $\A_\add(\V)$ commutes with $\A_\add(\U')$.    If one is hoping to make
  a general statement that would apply to any quantum field theory, this really forces 
  $\V\subset \U''$, since local operators outside of $\U''=(\U')'$ typically 
  do not commute with $\A_\add(\U')$.    So the most optimistic statement that one might hope for would be $\A_\add(\U)=\A_\add(\U'')$.
  
  For example, if $\U$ is a timelike curve in Minkowski space, then $\E(\U)=\U''$, and therefore the timelike tube theorem does indeed give $\A_\add(\U)=\A_\add(\U'')$.     
  In general, $\E(\U)\subset \U''$,
  but $\U''$ can be strictly bigger.   Indeed, in Araki's example, $\B''$ is strictly bigger than $\E(\B)$.   His conjecture would be a special case of $\A_\add(\U)=\A_\add(\U'')$, in a situation in which
  that does not follow directly from the known statement of the timelike tube theorem.
  
  A statement along the lines of $\A_\add(\U)=\A_\add(\U'')$ would be very attractive, as it would combine ordinary causality  with the timelike tube theorem.
  Extending from a spacelike pancake to its domain of dependence (fig. \ref{four}(a)) and from a timelike tube to its timelike envelope (fig. \ref{four}(b)) can be regarded
  as two different cases of extending from $\U$ to $\U''$.
  However, there are some easy counterexamples to  $\A_\add(\U)=\A_\add(\U'')$.   
  
\begin{figure}
 \begin{center}
   \includegraphics[width=1.3in]{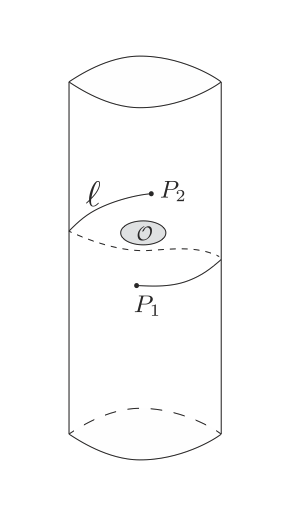}
 \end{center}
 \caption{\footnotesize In the spacetime $\R\times S^1$, with $\R$ parametrizing the ``time,'' $\ell$ is a spacelike curve that spirals around from a point $P_1$ to a point $P_2$
 that is  to its future.   $\O$ is an open set in the ``gap'' between $P_1$ and $P_2$.  If $\B$  (not drawn) is an open set that is a small ``thickening'' of $\ell$, then a right-moving
 null geodesic through $\O$ does not intersect $\B$.      \label{six}}
\end{figure}

  One type of counterexample\footnote{This example is relatively well known but the original reference is not clear.  A version of the example of fig. \ref{six} was discussed
  in \cite{Stroh}.}  
  arises if $\U$
  is not connected.    In Minkowski space in any even dimension $\sD$, let $\O$ be a small neighborhood of the origin. 
  Let $\U$ be any open set that contains points to the future of $\O$ and
  also points to the past of $\O$, but none that can be reached from $\O$ by a null geodesic.  Then $\O\subset \U''$, so $\A_\add(\U)=\A_\add(\U'')$ would 
  predict that $\A_\add(\O)\subset \A_\add(\U)$.  In the theory of a massless free  scalar field $\phi$,
  however, this statement is false.   By Huygen's principle along with commutativity at spacelike separation, the commutator function $G(x,y)= [\phi(x),\phi(y)]$ in that theory 
 is supported
  on the light cone (for $\sD$ even), so it vanishes for $x\in \O$, $y\in\U$.   Thus in this situation, $\A_\add(\O)$ commutes  with $\A_\add(\U)$, rather than being contained in $\A_\add(\U)$. 
  
  In two dimensions, a small variant of this gives a counterexample with $\U$ connected.   In fact,  the setup is closely related to that suggested by
  Araki.    Consider the spacetime $M=\R\times S^1$, with a product metric, where $\R$ parametrizes time and $S^1$ parametrizes space.   Let the point $P_2$ be slightly to the future
  of $P_1$, and let $\ell$ be a spacelike curve that starts at $P_1$, spirals once around $S^1$ to the right, and ends at $P_2$. Let $\B$ be an open set that is a slight thickening of
  $\ell$, and let $\Delta$ be the causal diamond with vertices $P_1,P_2$.  Finally, let $\O\subset \Delta$ be an open set small enough that any right-moving
  null geodesic that intersects $\O$ does not intersect $\B$  (fig \ref{six}).   Then $\O\subset \Delta\subset \B''$.
   In two-dimensional quantum field theory, it is possible to have a local field $J$ with the property that $[J(x),J(y)]$ vanishes unless
  the points $x$ and $y$ are connected by a right-moving null geodesic.   For example,  $J$ could be a right-moving conserved current (or a chiral component of the stress tensor)  in a conformally invariant theory.
Let $f$ be a smearing function supported in $\O$, and consider the operator $J_f=\int \d^2x f(x) J(x)\in \A_\add(\O)\subset\A_\add(\B'')$.   This operator commutes with $\A_\add(\B)$, rather than being
contained in $\A_\add(\B)$, as one would expect based on $\A_\add(\B)=\A_\add(\B'')$.

In the context of the present article, we are not primarily interested in applying the timelike tube theorem or a hypothetical generalization to an arbitrary open set.   
We are primarily interested in observables along the timelike worldline $\gamma$ of an observer.  The timelike tube theorem tells us that it does not matter whether we consider
a timelike curve $\gamma$ or an open set that is a suitable slight thickening of it, so we will express the following in terms of the algebra associated to a curve.
The example just discussed can be slightly
modified to give a counterexample  in that context.   If $P_2$ is farther to the future of $P_1$ than was assumed so far, then the  curve $\gamma$ that wraps around the $S^1$ 
 (in general any number of times) en route from $P_1$ to $P_2$
can be timelike, but almost null.   Then the timelike envelope $\E(\gamma)$ is  a small neighborhood of $\gamma$, but
its causal completion $\B''$, which is the same as $\gamma''$ if $\B$ is suitably chosen,  wraps all the way around the $S^1$.   The statement $\A_\add(\E(\gamma))=\A_\add(\gamma'')$ is false in general just as before.   

In this example, $\gamma''$ is the same as the causal diamond\footnote{Here $\Delta(\gamma)=J^+(\gamma)\cap J^-(\gamma)$ is the intersection of the past and future
of $\gamma$, or equivalently $J^-(P_2)\cap J^+(P_1)$.} $\Delta(\gamma)$ with vertices
 $P_1,P_2$, so in particular $\E(\gamma)\subsetneqq \Delta(\gamma)$.   Thus, this example shows that 
 in the statement of the timelike tube theorem, in general $\E(\gamma)$ cannot be replaced with\footnote{Similarly,  $\E(\gamma)\subsetneqq\Delta(\gamma)$ in a simply-connected spacetime $M=\R\times S^2$, with metric $\d s^2=-\d t^2+\d x^2+\d y^2+\epsilon \d z^2$, with $x^2+y^2+z^2=1$,
$\epsilon\ll 1$.   Embedding $\R\times S^1$ in $\R\times S^2$ at $z=0$ and choosing $\gamma$ as before, again $\Delta(\gamma)$ 
is much larger than $\E(\gamma)$ if $P_2$ is sufficiently far
to the future of $P_1$.   However, as in Araki's example, it is not clear in this case whether it is always true that $\A_\add(\gamma)=\A_\add(\Delta(\gamma))$.  Free field theory does
not provide an immediate counterexample.} 
$\Delta(\gamma)$.

So some simple generalizations of the timelike tube theorem are false in a general quantum field theory, and others, such as  Araki's conjecture, have unclear status.
However, the following somewhat fanciful remarks come to mind.    Let us go back to the example in which $\U$ is  not connected and is, for example, 
the union of a small ball $\B_q$ around a point $q$
and a small ball $\B_p$ around a point $p$ to its future. (Similar remarks apply in the other examples.)
In this case, for $\sD$ even, massless free field theory provides a counterexample to $\A_\add(\U)=\A_\add(\U'')$, but the argument was limited
to massless free field theory and would not generalize in an obvious way to any other theory.    Consider then in a more generic theory an observer\footnote{In this paragraph, unlike the rest of the article, we consider an
observer probing the spacetime from outside, not an observer who propagates on a worldline in the spacetime.}  who has the capability to
manipulate the quantum fields in an arbitrary fashion, but only in $\U=\B_p\cup \B_q$.   A strategy this observer can follow is to inject into spacetime a probe in the region $\B_q$,
directed on a trajectory $\gamma$ that will carry it to $\B_p$, and equipped and programmed to make some chosen measurement 
and  record the result.   Then the observer retrieves the probe at $\B_p$ and reads the answer.   Does this construction along with the timelike tube theorem
prove that $\A_\add(\E(\gamma))\subset \A_\add(\U)$? 
  Though we assume no particular limit on the technological capabilities of the observer, we assume that whatever happens in a part of spacetime to which the 
  observer does not have direct access is governed by the theory that is being probed.  Thus in particular the probe that the observer injects into spacetime in $\B_q$ must be something that can be built
 in this theory.   A complex probe cannot be built  in free field theory, so there is no tension between existence of this protocol and our earlier observations about free field theory.
  However, in a sufficiently complex theory -- possibly any theory  that encompasses the Standard Model, for example -- this protocol does indeed hint
   that $\A_\add(\E(\gamma))\subset \A_\add(\U)$.
Actually, the observer can choose $\gamma$ at will (assuming that the probe can be equipped with a rocket engine and programmed to travel on a pre-chosen trajectory),
  and  more generally the observer could inject into the system several probes with pre-chosen trajectories $\gamma_i$ from $\B(q)$ to $\B(p)$, each designed and programmed to carry out a 
  particular quantum operation, and then the observer can collect the probes in $\B(p)$ and process their (possibly quantum) output at will.   All this suggests that in a theory that is sufficiently complex, possibly in the end $\A_\add(\U)=\A_\add(\U'')$ in general.

 \subsection{Causal Wedge Reconstruction}\label{causalwedge}
 
 There is actually a manifestation of the timelike tube theorem that has been much discussed in the literature.   This is causal wedge reconstruction, or HKLL reconstruction,
 in the context of the AdS/CFT correspondence  \cite{BDHM,B,Bal,HKLL,HKLL2,KLL,HMPS,Mor,H}.
 
 In the AdS/CFT correspondence, one considers a conformal field theory on a globally hyperbolic spacetime $M$ of dimension $\sD$.   AdS/CFT duality says that
 an appropriate conformal field theory on $M$ is equivalent to a gravitational theory formulated on a spacetime $X$ that has $M$ for its conformal boundary, and that is globally
 hyperbolic in the asymptotically AdS sense.   In fact, in AdS/CFT duality, one has to consider all possible $X$'s that have a given conformal boundary $M$, but for our
 purposes here, we can assume that a particular $X$ is important.   Causal wedge reconstruction applies in a semiclassical situation in which
 $X$ can be viewed, in leading order, as a definite spacetime in which quantum fields are propagating.

  \begin{figure}
 \begin{center}
   \includegraphics[width=2in]{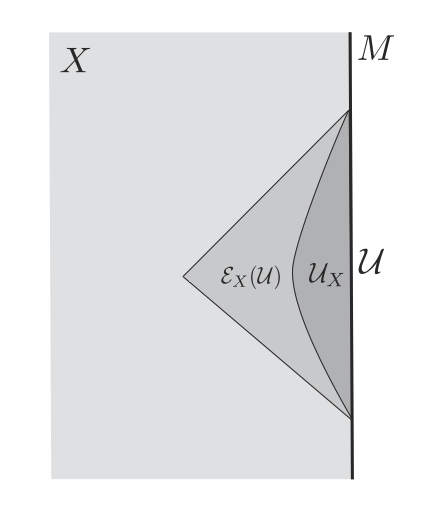}
 \end{center}
\caption{\footnotesize A region $\U$ in the boundary $M$ of $X$, and a slight thickening $\U_X$ of $\U$ in the bulk spacetime $X$, chosen so that
 $\E_X(\U)=\E(\U_X)$. \label{seven}}
\end{figure} 
 
 Let $S$ be a Cauchy hypersurface in $M$, $B$ an open set in $S$, and $\U$ the domain of dependence of $B$ in $M$.    Then $\U$ is its own timelike envelope in $M$, so the timelike
 tube theorem, applied directly to the region $\U$ in the CFT on $M$, says nothing of interest.  However, if $M$ is the conformal boundary of $X$ and $\U\subset M$,
 then it makes sense to consider the timelike envelope $\E_X(\U)$ of $\U$  in $X$ (fig. \ref{seven}).
In its simplest form, causal wedge reconstruction then asserts that
 $\A_\add(\U)=\A_\add(\E_X(\U))$.    This can be viewed as a composite of two statements: (1) the basic AdS/CFT duality expressing local operators in $M$ as limits of local
 operators in $X$, and (2) the timelike tube theorem applied to quantum fields in $X$.
 
 To explain this, we can proceed as follows.   First, it is possible to thicken $\U$ slightly to an open set $\U_X\subset X$ such that $\E_X(\U)=\E(\U_X)$.   
The timelike tube theorem says that $\A_\add(\U_X)$ does not depend on the precise choice of $\U_X$, and always equals $\A_\add(\E(\U_X))$.   Since
this is the case, we can take a limit in which $\U_X$ becomes  arbitrarily ``thin.''   The basic AdS/CFT relation 
between local operators on $X$ and local operators on $M$ can be interpreted as a statement
\be\label{zono} \lim_{\U_X\downarrow\U}\A_\add(\U_X)=\A_\add(\U) ,\ee
where $\lim_{\U_X\downarrow\U}$ refers to a limit in which the bulk open set $\U_X$ collapses down to the boundary open set $\U$.  On the left hand side, $\A_\add(\U_X)$ is the algebra
generated by bulk local operators in the bulk region $\U_X$, and on the right hand side, $\A_\add(\U)$ is the algebra generated by  boundary local operators in the boundary region $\U$.
Eqn. (\ref{zono}) is a way to express the fact that boundary local operators are the boundary limits of bulk local operators.
Since the timelike tube theorem tells us that $\A_\add(\U_X)=\A_\add(\E(\U_X))=\A_\add(\E_X(\U))$ regardless of the choice of $\U_X$, 
the limit in eqn. (\ref{zono}) is trivial and we learn that
\be\label{bono} \A_\add(\E_X(\U)) =\A_\add(\U).\ee

In a globally hyperbolic spacetime that satisfies the null energy condition,
$\E_X(\U)$ is the causal wedge of $\U$, and eqn. (\ref{bono}) is  the usual statement of causal wedge reconstruction.   However, this requires some explanation,\footnote{The following
proof is not needed in the rest of the article. 
Facts used in the  proof are explained, for example, in \cite{Wald,GaoWald,LightRays}.  On
compactness of spaces of causal curves, see for instance sections 2 and 3.3 of \cite{LightRays};
on the fact that a causal curve that satisfies a promptness condition is a null geodesic without focal points, see sections 5.1 and 5.2 of that article; on the completeness of  a null geodesic
in an asymptotically AdS spacetime $X$
whose ends are on the conformal boundary of $X$, see section 7.3; on the fact that a complete null geodesic along which the null energy condition is satisfied as a strict inequality must have
focal points, see section 8.2.}
since the usual definition of the causal wedge is slightly different.    The causal wedge of $\U$, which we will denote as $\CC(\U)$, is usually defined to consist of all points
$r\in X$ that are contained in a causal curve  $\gamma\subset X$ between two points $q,p\in \U$, say with $p$ to the future of $q$.   (Equivalently, $\CC(\U) $ is the intersection
of the past and future of $\U$ in $X$.)
Instead, $\E_X(\U)$ contains all points $r\in X$ that are contained in a causal curve
$\gamma\subset X$ between points $q,p\in \U$ such that $\gamma$ can be deformed, through a family of causal curves with fixed endpoints, 
to a causal curve entirely in $M$ (and therefore in $\U$).   Thus to show
that $\E_X(\U)=\CC(\U)$, we have to show that any causal curve $\gamma\subset X$ that starts and ends at points $q,p\in \U$ can be deformed, through a family of causal curves
with those fixed endpoints, to a causal curve in $M$.   To prove this, let $\RR_{qp}$ be the space of causal curves in $X$ with initial and final endpoints $q,p$.
 Pick a causal curve $\zeta_{qp}\subset \U $ from $q$ to $p$, and let $\TT_{\zeta}$ be the space of causal curves in $X$
 with initial endpoint $q$ and any final endpoint $s\in \zeta_{qp}$.    A causal curve from $q$ to $s$ can be converted to a causal curve from $q$ to $p$ by gluing on the
 segment $\zeta_{sp}$ of $\zeta_{qp}$.
 This operation maps $\TT_\zeta$ continuously to $\RR_{qp}$; on the other hand, $\RR_{qp}\subset \TT_\zeta$.
So a causal curve $\gamma\subset X$ with endpoints $q,p$ can be deformed in $\RR_{qp}$ to a curve entirely in $M$ if and only if it can be so deformed in $\TT_\zeta$.
Hence, to prove that $\E_X(\U)=\CC(\U)$, it suffices to show that any causal curve $\gamma\subset X$ with endpoints $q,p$  can be deformed in $\TT_\zeta$ to a curve entirely in $M$.
This question is topological in nature: is $\gamma$ in the connected component of $\TT_\zeta$ that contains curves that lie entirely in $M$?   So the answer is invariant under
an infinitesimal perturbation of $X$, and we can make such a perturbation to ensure that the null energy condition is satisfied in $X$ as a strict inequality (not as an equality).   
We will see later exactly where in $X$ the perturbation should be made.
On $\TT_\zeta$, one can define the following continuous function $\varphi$: if $\gamma_*\in\TT_\gamma$ has endpoints $q,s\in\zeta_{qp}$, then $\varphi(\gamma_*)$ is the
proper time elapsed from $q$ to $s$ along $\zeta_{qp}$.   An important detail is that we allow the case of a causal curve that consists of 
only one point.  Thus in particular
$\TT_\zeta$ contains a point $q_*$ that corresponds to a causal curve consisting only of the point $q$.   
The point $q_*$ is the absolute minimum of $\varphi$,  since $\varphi(q_*)=0$, and at other points in $\TT_\zeta$, 
$\varphi>0$.  The purpose of including the point $q_*$ in the definition of $\TT_\zeta$ is to ensure that $\TT_\zeta$ is compact; this compactness follows
from the fact that the initial endpoint of a curve $\gamma_*\in \TT_\zeta$ is fixed and its final endpoint $s$  ranges over the compact set $\zeta_{qp}$, along with the fact that in a globally hyperbolic spacetime, the space of causal curves with
specified endpoints is compact.     Compactness of $\TT_\zeta$ ensures that in every connected component of $\TT_\zeta$, the nonnegative function $\varphi$ has an absolute
minimum. Any  $\gamma\in\TT_\zeta$ can be deformed in $\TT_\zeta$ to the minimum of $\varphi$ in its connected component.   
We will conclude the proof by showing that the point $q_*$ is the unique local minimum of the function $\varphi$, implying that the local minimum 
of $\varphi$ to which $\gamma$ can be deformed 
in $\TT_\zeta$ is actually the absolute minimum $q_*$.   This in particular implies that $\gamma$ can be deformed in $\TT_\zeta$ to a causal curve entirely in 
$M$, as we wished to show.
To show that  $\varphi$ has  no other local minimum, we note that any point in $\TT_\zeta$ other than $q_*$ is a nontrivial causal curve 
$\t\gamma$ with distinct endpoints in $M$.   In general, if the initial point of a causal curve $\t\gamma$ is specified  (here $q$)  and the final point of the curve is constrained by
a condition of ``promptness'' (here $\t\gamma$ is supposed to end along $\zeta_{qp}$, and the condition for it to be a local minimum of $\varphi$ means that it arrives on $\zeta_{qp}$ sooner than any nearby 
causal curve from $q$; this is the promptness condition),  then $\t\gamma$ must be a null geodesic without focal points.  
However, in an asymptotically AdS
spacetime, any null geodesic between distinct points on the conformal boundary is complete in both directions.   As remarked earlier,
 we can assume that the null energy condition in $X$ is satisfied along
$\t\gamma$
as a strict inequality.   A null geodesic that is complete in both directions and along which the null energy condition is satisfied as a strict inequality always has focal points.   
So the function $\varphi$ can have no local minimum other than its absolute minimum $q_*$, completing the proof.

\section{An Algebra of Observables For De Sitter Space}\label{desitter}

Finally, we turn to a concrete example in which,  in order to define a sensible algebra of observables,
it is necessary to include an observer in the description  \cite{CLPW}.
De Sitter space in $\sD$ dimensions or $\dS_\sD$  is the maximally symmetric solution of Einstein's
equations with a positive cosmological constant.     It can be described by the metric
\be\label{telfo}\d s^2 = -\d t^2+R^2 \cosh^2(t/R) \d\Omega^2,\ee
where $R$ is the radius of curvature and 
$\d\Omega^2$ is the metric of a round sphere $S^{\sD-1}$ of unit radius.   This sphere is compact, so $\dS_\sD$ is an example of a        closed universe.  At time $t$, the sphere has radius
$R(t) = R \cosh(t/R),$
so it grows exponentially  for $t\to +\infty$ or $t\to -\infty$.     The exponential growth for $t\to+\infty$ is believed to be a good approximation to what is currently beginning to
 happen in the real world.
 
 In the 1970's, Gibbons and Hawking \cite{GH} studied de Sitter space as a simple example of a spacetime with a cosmological horizon -- in which an observer cannot see
the whole universe.   They attached a temperature and  entropy to the de Sitter horizon, as Bekenstein and Hawking had done not long before for the horizon of  a black hole.  The
thermal interpretation is most obvious in Euclidean signature, where $\dS_\sD$ becomes simply a $\sD$-sphere $S^\sD$, with metric
  \be\label{sphere}\d s^2_E=\d\tau^2+R^2\cos^2(\tau/R)\d\Omega^2.\ee
In ordinary quantum field theory in de Sitter space (and also in the presence of semiclassical gravity), there is a natural de Sitter state $\Psi_\dS$
  such that correlation functions in this state can be obtained by analytic continuation from Euclidean signature.   Let $C$ be a great circle in $S^{\sD}$ and let
  $\lambda$ be a length parameter along $C$.    $C$ is an orbit of a  $\UU(1)$ symmetry of $S^\sD$ (which is uniquely determined if we say that its fixed point set is a copy
  of $S^{\sD-2}$).   If we normalize the generator $H$ of this
  $\UU(1)$ to act as $\partial/\partial \lambda$ along $C$,  then $H$ obeys  $\exp(-2\pi R H)=1$.
When continued to Lorentz signature, this leads to the striking statement  that correlation  functions in the state $\Psi_\dS$ have a thermal interpretation
  at the de Sitter temperature $T_\dS = 1/\beta_\dS$, where $\beta_\dS=2\pi R$ \cite{GH,FHN}.    The thermal interpretation of de Sitter space has been extensively explored
  for nearly half a century.   For a small sampling of the relevant literature, see 
\cite{Sewell,Maeda,BoussoOne,BoussoTwo,Banks,BanksFischler,BanksOne,BanksTwo,BFTwo,SusskindA,Susskind,DF,BD,SB,DST}.

      \begin{figure}
 \begin{center}
   \includegraphics[width=2.6in]{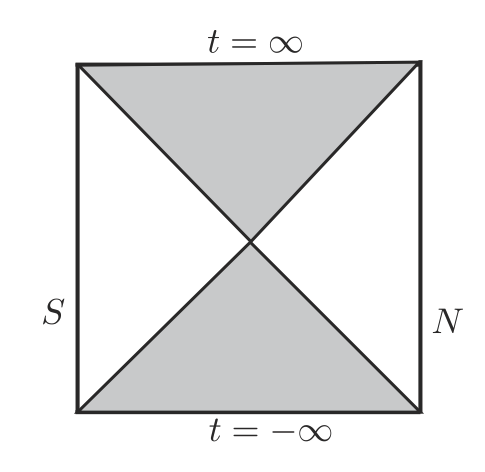}
 \end{center}
 \caption{\footnotesize The Penrose diagram of de Sitter space.  Past infinity is at the bottom; future infinity is at the top.
 The left and right boundaries represent timelike geodesics that can be viewed as the trajectories of  observers
 who remains at rest on the south or north pole of $S^{\sD-1}$, respectively.   
The diagonals represent the past and future horizons of those observers; the unshaded triangles on the left and right 
are the static patches
that are causally accessible to these observers. \label{eight}}
\end{figure}  
  
In Lorentz signature, de Sitter space is conveniently understood via   a Penrose
diagram (fig. \ref{eight}).  The great circle $C$ continues in Lorentz signature to a hyperbola that has two components, each of them a geodesic.
Coordinates can be chosen so that these geodesics make up the left and right boundaries of the Penrose diagram.
Given an observer traveling on a geodesic in $\dS_\sD$, 
we can assume that  the worldline of this observer is, say, the left boundary of the diagram.
The observer then has past and future horizons which are the diagonals in the picture.   The region causally accessible to the observer  (the region the observer can see and
also can influence) 
is bounded by these diagonals, along with the left boundary of the diagram.    A similar region of the diagram is causally accessible to an observer
on the right boundary.   The operator  $H$ generates a symmetry of the Penrose diagram; with a suitable choice of sign,
  it maps the region accessible to the left observer  forward in time and the region accessible to the right observer  backwards in time.  
  Near the bifurcate horizon where the two diagonals meet, $H$ looks like the generator of a Lorentz boost.
     
  If $t$ measures the time along the left boundary, then on the left boundary of the figure, $H=\partial/\partial t$.  So it is natural for an observer whose worldline is the left
  boundary to interpret $H$ as a generator of time translations.    With this interpretation, since $H$ is a symmetry of the region causally accessible to the left observer,
  this region is 
 time-independent and thus ``static.''   This is why the region causally accessible to an observer in de Sitter space has been called a ``static patch.''
However,  this view of de Sitter space as being ``static'' is
  highly observer-dependent.   In a global view, the form (\ref{telfo}) of the de Sitter metric shows that $\dS_\sD$ is expanding exponentially both toward the future and toward the past,
  so globally one would definitely not call $\dS_\sD$ ``static.'' 
   
An ordinary quantum field theory in de Sitter space has a Hilbert space $\H_0$ of quantum states.   In such a theory, 
  we associate to the static patch (or any region) an algebra $\A_0$ of observables consisting of operators on $\H_0$ that act on the quantum fields in the region in question.   
The algebra   of any local region, and in particular the algebra $\A_0$ of the static patch, is a possibly unfamiliar Type III von Neumann algebra. 
  This is an algebra with an infinite amount of quantum entanglement built in, giving an abstract explanation of the fact that entanglement entropy is ultraviolet divergent in quantum
  field theory.   Including weakly coupled gravitational 
  fluctuations does not qualitatively change the picture.   We simply include the weakly coupled gravitational field as one more field in the construction of $\H_0$ and $\A_0$.
     What does really change the picture is that in a  closed universe, such as de Sitter space,
   the isometries have to be treated as constraints.  In the case of the static patch, the important constraint is the Hamiltonian $H$.    Imposing $H$ as a constraint means
    that we should replace $\A_0$ by
  $\A_0^H$, its invariant subalgebra.    But that does not work:   the invariant subalgebra is trivial.      Roughly, that is because anything that commutes with $H$
  can be averaged over all the thermal fluctuations and replaced by its thermal average, a $c$-number.  A technical statement is that there are no nontrivial invariants in $\A_0$ because
  $H$ generates the modular automorphism group of the state $\Psi_\dS$ for the algebra $\A_0$ \cite{Sewell}, and therefore acts ergodically.
  
  To get a reasonable algebra of observables, we include an observer in the analysis.    Of course, as noted in the introduction, in principle an observer should really be
  described by the theory,
  not  injected from outside.    What it means to include an observer is that we consider a ``code subspace'' of states in which an observer is
  present in the static patch, and then we consider operators that can be defined in the low energy effective field theory in this code subspace, though they 
  are not well-defined on the whole Hilbert space.
  
  Should we be surprised that we need to include the observer in the analysis to get a sensible answer?    As was also noted in the introduction, a gravitating system in a closed
  universe is the situation in which we are most likely to need to explicitly incorporate the observer in the analysis.   That is exactly the situation  here
  because de Sitter space is a simple model of a closed universe, that is, a universe with compact spatial sections.   
    
  Once we include an observer in the analysis, there is a rationale to study a particular static patch, namely  the region that is causally accessible to that given observer.
Moreover, the timelike tube theorem tells us that the algebra $\A_0$ of observables in the causally accessible region can be interpreted as 
  the algebra generated by the quantum fields along the observer's worldline.   Thus the algebra $\A_0$ of the static patch becomes, in a sense, operationally meaningful
  once the observer is included.  
  
It turns out that to get a sensible answer, it suffices to consider a minimal model in which the observer is characterized just by a clock with Hamiltonian
  \be\label{hobs} H_{\mathrm{obs}}=q.\ee
     It is physically reasonable to assume that the observer's energy is bounded below.  The precise lower bound will not be important and we will just take it\footnote{More realistically, the observer would have a mass $m$ and minimum energy $mc^2$.  If the observer is minimally coupled to gravity, then the observer worldline will
     be a geodesic.   In the presence of an observer, the relevant static patch is the region causally accessible to
     the observer.} to be 0.
    With that choice, the effect of including the observer is to modify the Hilbert space by
  \be\label{modhilb} \H_0\to \H_1=\H_0\otimes L^2(\R_+), \ee
where $q$ is a multiplication operator on the positive half-line $\R_+$.     The algebra is likewise extended from $\A_0$ to
\be\label{modalg}\A_1=\A_0\otimes B(L^2(\R_+)).\ee
The  last factor is the (Type I) algebra of all bounded operators on $L^2(\R_+)$.

Finally the constraint becomes the total Hamiltonian of the quantum fields plus the observer:
\be\label{nobs}H\to \h H=H+H_{\mathrm{obs}}.  \ee   The ``correct'' algebra of observables taking account of the presence of the observer  is therefore
\be\label{thalg} \A=\A_1^{\h H}, \ee
that is, the $\h H$-invariant part of $\A_1$.    To be more exact,  this is the algebra of observables accessible to the observer in the limit $G_N\to 0$.
In higher order in $G_N$, it will be necessary to incorporate direct couplings between the observer and the quantum fields; the constraint operator will not be a simple
sum.   Of course, the notion of the algebra of a spacetime region is  presumably only well-defined in the limit $G_N\to 0$;   the timelike tube theorem suggests that in going beyond that
limit, we should reinterpret
$\A$ as the algebra of observables along the observer's worldline.   As explained presently, the
 important conclusions that we will draw about $\A$ are expected to be robust against perturbative
corrections in $G_N$. 

The  answer (\ref{thalg}) makes sense, unlike the previous one.    The reason is that once an observer is present, we can ``gravitationally dress''  any operator to the observer's world-line.   
It is easiest to explain this if we momentarily ignore the lower bound $q\geq 0$.   In fact, in the field of a black hole, there is a problem that is mathematically quite similar
to what we are discussing here, but without the lower bound on $q$ \cite{CLPW}.   
If we do ignore the lower bound on $q$, then the Hilbert space including the observer is $\H_1'=\H_0\otimes L^2(\R)$,
and the algebra is $\A_1'=\A_0\otimes B(L^2(\R))$.   Here $B(L^2(\R))$ is generated by (bounded functions of) $q$ and $p=-\i \frac{\d}{\d q}$.   If $\Theta$ is the projection 
operator onto states with $q\geq 0$, then $\H_1=\Theta \H_1'$, $\A_1=\Theta \A_1'\Theta$.   Similarly the $\h H$-invariant subalgebras $\A=\A_1^{\h H}$, $\A'=\A_1'{}^{\h H}$ are related
by $\A=\Theta \A'\Theta$.   So $\A$ is easily constructed once we understand $\A'$.

To construct $\A'$, we reason as follows.
  For any $\a\in \A_0$, the operator
 $$\h\a=e^{\i p H}\a e^{-\i p H}$$ commutes with the constraint $\h H=H+q$.      One more operator that commutes with the constraint is $q$ itself (or equivalently
 $-H$, which equals $q$ modulo the constraint).      It follows from a classic result of Takesaki \cite{Takesaki} that by coincidence was proved in the early days of black
 hole thermodynamics 
  that (1) there are no more operators in $\A_1'$ that commute with the constraint,
 and (2) the algebra $ \A'$ that is generated by the operators $\h \a$ and $q$ is actually a von Neumann algebra of Type II$_\infty$, with a trivial center (that is, its center consists only
 of $c$-numbers).   In the case of the static
 patch in de Sitter space, we want to impose the constraint $q\geq 0$, so the appropriate algebra is not $\A'$ but $\A=\Theta \A\Theta$.   This is an algebra of Type II$_1$, again with
 trivial center.

A basic introduction to the relevant facts about von Neumann algebras  can be found in \cite{CLPW}, among other places.  For much more depth, see \cite{Sorce}. 
An important fact  for our purposes is that a Type II
algebra $\A$  (unlike one of Type III, which we would have in the absence of gravity) has a trace, that is a  complex-valued linear function obeying $\Tr\,\a\b=\Tr\,\b\a$, for all $\a,\b\in\A$.  
Moreover, the trace in a Type II  algebra is positive, in the sense that $\Tr\,\a\a^\dagger>0$ for all $\a\in\A$.

A von Neumann algebra with trivial center is called a factor.  Thus the above analysis implies that in the limit $G_N=0$, the algebra of observables in the static
patch is a factor of Type II$_1$.    A factor is the von Neumann algebra analog of a simple Lie group.   A simple Lie group  is rigid, in the sense that it has has no infinitesimal deformations.
That is not true for a non-simple Lie group; for example, the symmetry group of $\R^n$ is non-simple, and can be deformed, by an arbitrarily small perturbation of the commutation relations
in its Lie algebra, to the symmetry group $\SO(n+1)$ of $S^n$.   A similar statement holds for von Neumann algebras: an algebra with a non-trivial center can potentially be deformed 
to an algebra of a different type, while making the center smaller,\footnote{For example, the Type III algebra described by Leutheusser and Liu in the $G_N\to 0$ limit
in the field of a black hole \cite{LL,LLtwo}  has a nontrivial center and is modified by perturbative corrections to a factor of Type II$_\infty$ \cite{CLPW,Witt}.}
 but a factor is rigid and has no infinitesimal deformations.   
 Though we have only analyzed the algebra of observables in the static patch in the limit $G_N\to 0$, because the answer that we obtained is a factor,
perturbative corrections in $G_N$ are not expected to modify the algebra up to isomorphism (they will modify the commutation relations among operators along
the observer's worldline, but not the isomorphism class of the algebra that those operators generate).   
Nonperturbatively, matters are unclear and indeed it is quite unclear whether quantum de Sitter space
makes sense nonperturbatively.   If quantum de Sitter space does make sense nonperturbatively, then one expects to describe it by a finite-dimensional Hilbert space  \cite{BanksOne,BanksTwo},
and the algebra will have to be of Type I.

A Type II algebra does not have pure states, but because it does have a trace, 
  familiar ideas like density matrices and entropies make sense for a state of such an algebra \cite{Segal,LW}.    To a global state $\Psi$ of de Sitter space plus
 the observer,  reduced to the static patch, one can associate
 a density matrix $\rho\in\A$, characterized by
 \be\label{zogg} \la\Psi|\a|\Psi\ra =\Tr\,\a\rho,~~\forall \a\in\A.\ee     Therefore, we can define a von Neumann entropy
 \be\label{vonent} S(\rho)=-\Tr\,\rho\log \rho.\ee  
 There is no such definition in the absence of gravity, because without gravity, the observables in the static patch constitute the Type III algebra $\A_0$.
   The fact that gravity turns the Type III algebra into a Type II algebra gives an abstract explanation for why entropy is better defined in the presence of gravity
 than in ordinary quantum field theory.   However,  from a physical point of view, Type II entropy is a renormalized entropy from which an
 infinite constant has been subtracted.

As already remarked, if we put no constraint on the observer's energy, we get an algebra of Type II$_\infty$; if we assume the observer's energy is bounded below, 
we get an algebra of Type II$_1$.  In a Type II$_1$ algebra, the trace is defined for all elements of the algebra,
 while in an algebra of Type II$_\infty$,
the trace is finite only for a dense subset of elements of the algebra.   In particular, for Type II$_1$,  the identity element has a finite trace and we can normalize the trace so that the trace
of the identity is 1,
while in Type II$_\infty$, the trace of the identity element is $+\infty$.

For our purposes, the important difference is that a Type II$_1$ algebra has  a state of maximum possible entropy, which is the ``maximally mixed'' state with density matrix $\rho=1$.    
 This is consistent with $\Tr\,\rho=1$, assuming that the trace has been appropriately normalized.
 Evaluating $S(\rho)=-\Tr\,\rho\log\rho$ for $\rho=1$, we see that the state with $\rho=1$ has entropy 0.   It is not difficult to prove that all other states have negative entropy (for example,
 see \cite{CLPW,LW}).   This tells us the  meaning of the subtraction that is involved in defining the entropy of a state of a
 Type II$_1$  algebra so that $S(\rho)=0$ for $\rho=1$: the constant that is subtracted is the maximum possible entropy. In general, entropy of a state of a Type
 II$_1$ algebra  is the entropy difference of that state 
  relative to the entropy of the maximum entropy state.     The maximum entropy state with $\rho=1$ is the Type II$_1$ analog of a maximally 
 mixed state in ordinary quantum mechanics, in which the density matrix
is a multiple of the identity.  By contrast, there is no upper bound on the entropy of a state of a Type II$_\infty$ algebra.

It is felicitous that  a Type II$_1$ algebra has a state of maximum entropy,   because in fact 
de Sitter space is believed to have a state of maximum entropy -- namely ``empty de Sitter space,'' with all the entropy in the cosmological horizon \cite{Maeda,BoussoOne,BoussoTwo}.
  So to get a reasonable model of de Sitter space, it is important to  assume that the observer's energy is bounded below, which is a more reasonable assumption anyway, thereby 
  ensuring that the algebra of observables is of Type II$_1$, not Type II$_\infty$. 
  One can explicitly construct a  state $\Psi_\max\in \H_\dS$ that has maximum entropy (and thus density matrix $\rho=1$) 
  when reduced to the static patch.  In fact, one can pick $\Psi_\max =\Psi_\dS\otimes e^{-\beta_{\dS} q/2}\sqrt{\beta_\dS}$.   Thus empty de Sitter space, 
  tensored with a  state in which the observer's energy has a thermal distribution at the de Sitter temperature, has maximum entropy from the point of view of the Type II$_1$ algebra.
In this sense the analysis based on the Type II$_1$ algebra agrees with the claim that empty de Sitter space has maximum entropy.

We can now compare with some further claims in the previous literature.    First of all, since the maximum
entropy state has $\rho=1$, it has a ``flat entanglement spectrum'' (all eigenvalues of the density matrix are equal) and accordingly the R\'{e}nyi entropies all vanish:
\be\label{renyi} S_\alpha(\rho)=\frac{1}{1-\alpha}\log \,\Tr\,\rho^\alpha = 0. \ee
This matches with a result that has been found
using Euclidean path integrals to analyze the R\'{e}nyi entropies of the static patch \cite{DST}.
  Given the assertion that de Sitter space has a state of maximum entropy, this  result is what one should expect.
 In ordinary quantum mechanics, the maximum entropy state of a system is ``maximally mixed,'' with a ``flat entanglement spectrum'' (the density matrix
is a multiple of the identity and all its eigenvalues are equal) and its R\'{e}nyi entropies are independent of $\alpha$.   

Now, suppose that the observer makes a measurement with two outcomes that correspond  to the projection operators $\Pi$ and $1-\Pi$.    The probabilities of the
two outcomes are $\Tr\,\Pi$ and $\Tr\,(1-\Pi)=1-\Tr\,\Pi$.    All values $0\leq \Tr\,\Pi\leq 1$ are possible.    If the outcome corresponding to $\Pi$ is observed,
then after this measurement, the density matrix is
$$\sigma=\frac{1}{\Tr\,\Pi}\Pi. $$
Since the two eigenvalues of $\sigma$ are 0 and $1/\Tr\,\Pi$, one has $\sigma\log\sigma =\sigma \log (1/\Tr\,\Pi)$ so the entropy after the observation is
$$S(\sigma)=-\Tr\,\sigma\log\sigma = -\log(1/\Tr\,\Pi).$$
The entropy reduction from knowing the outcome is therefore $\Delta S = \log(1/\Tr\,\Pi)$, and this is related to the probability $p=\Tr\,\Pi$ of the given outcome by 
$$p = e^{-\Delta S}.$$

However, the probability of a (low entropy)  energy $E$ fluctuation of the static patch is
$$p= e^{-\beta_\dS E},$$
according to the thermal interpretation of de Sitter space.        Since also $p=e^{-\Delta S}$, we must have for consistency of the two descriptions
$$e^{-\beta_\dS E}= e^{-\Delta S}.$$
  In other words,  ```thermal'' suppression of a fluctuation can be understood as purely entropic suppression.   This is surprising, but it has been argued before on other grounds, notably by considering the case that the ``fluctuation'' is a small black hole at the center of the static patch \cite{Susskind}.

Which part of this is unexpected?  The 
 formula $p=e^{-\Delta S}$ for the probability of an outcome is an inevitable consequence of having a maximum entropy state in which all states
are equally probable.    In other words, if all states are equally likely, then the probability of a given outcome is just proportional to the number of microstates
that are compatible with that outcome.    Here we use language appropriate for an ordinary quantum system with a finite-dimensional Hilbert space.
For a system described by a  Type II$_1$ algebra, the number of microstates compatible with any given outcome is infinite, and one has to express the argument in terms of traces,
as we did earlier.  In short, the surprise is not that $p=e^{-\Delta S}$, which one should expect for a maximum entropy state, but that after coupling to gravity and including the observer,
the thermal state $\Psi_\dS$ can be promoted to a maximal entropy state $\Psi_\max$.

Let us see explicitly how this happens at the level of correlation functions.\footnote{An error in a previous claim about this matter  was pointed out by G. Penington.}
First we recall some basic facts about time-dependent correlation functions in a thermal ensemble.   The 
time dependence of an operator is defined in the usual way by  $\a(t)=e^{\i Ht}\a e^{-\i H t}$.   A typical time-dependent two-point function is 
\be\label{thermal} F(t)=\la \a(t) \b(0) \ra_\beta =\frac{1}{Z}\tr\,e^{-\beta H}\a(t)\b(0)=\frac{1}{Z}\tr\,e^{-\beta H}e^{\i H t} \a e^{-\i H t} \b.\ee  Here
$\tr$ is the trace in the Hilbert space of a thermal system with Hamiltonian $H$, and $Z$ is the partition function.
It follows immediately from the definition that $F(t)$ is holomorphic in a strip $0\geq \,\Im\,t\geq -\beta$ and moreover that
the boundary value at $\Im\,t=\-\beta$ is the thermal correlator with the opposite ordering of the operators.
In other words, let
\be\label{upthermal} G(t)=\la \b(0)\a(t)\ra_\beta =\frac{1}{Z}\tr\,e^{-\beta H} \b e^{\i H t}\a e^{-\i H t}. \ee
This function is related to $F$ by
\be\label{relation} G(t)=F(t-\i \beta). \ee
The precise meaning of this statement is that there is a function holomorphic in the strip $0\geq \Im\,t\geq -\beta$
whose boundary value on the upper boundary is $F$, while its boundary value on the lower boundary is $G$.

Let us express this in terms of Fourier transforms of the two functions.  Suppose that
\begin{align}\label{fourier} F(t)&=\int_{-\infty}^\infty \frac{\d \omega}{2\pi} e^{-\i \omega t} f(\omega)
\cr G(t)&=\int_{-\infty}^\infty \frac{\d \omega}{2\pi} e^{-\i \omega t} g(\omega).
\end{align}
Then eqn. (\ref{relation}) becomes
\be\label{undu} g(\omega)= e^{-\beta \omega} f(\omega).\ee
 
These facts about an ordinary quantum system also hold in de Sitter space
 for time-dependent correlation functions in the state $\Psi_\dS$, though  more sophisticated proofs are required.   To  make precisely the same argument for thermal correlators in the static patch of de Sitter space that
 one can make for an ordinary thermal system, one should have a Hilbert space
that describes the static patch, and $\tr$ should be the trace in this Hilbert space.   Such a ``one-sided'' Hilbert space does not exist for quantum fields in de Sitter space, because
the algebra of the static patch is of Type III, not Type I.  Instead one
only has a global Hilbert space describing all of de Sitter space.   To deal with this situation requires more careful arguments using Tomita-Takesaki theory \cite{Sewell}.   

What happens after coupling to gravity? As described earlier, we introduce an observer with energy $q$, and canonical momentum $p=-\i \d/\d q$. Let $\Theta(x)$
be the function that is 1 for $x\geq 0$ and vanishes for $x<0$, so that  $\Theta=\Theta(q)$ is the projection operator
onto states of $q\geq 0$.  To get a model of de Sitter space with weakly coupled gravity, we replace operators $\a$, $\b$ of the original thermal system by gravitationally dressed 
versions
\be\label{dressed}\h \a =\Theta e^{\i p H}\a e^{-\i p H}\Theta, ~~~\h\b =\Theta e^{\i p H}\b e^{-\i p H}\Theta.  \ee  
If we do not include the projection operators $\Theta$, the algebra would be of Type II$_\infty$, with no maximum entropy state.    Including the projection operators gives
a Type II$_1$ algebra with a maximum entropy state. 
As claimed earlier, this state is 
\be\label{psimax}\Psi_\max =\Psi_\dS \otimes\Psi_\beta(q) ,\ee
where $\Psi_\dS$ is the natural de Sitter invariant state in the absence of gravity, and
\be\label{psibeta}\Psi_\beta(q) =\begin{cases} e^{-\beta q/2}\sqrt \beta  &q\geq 0 \cr 0& q<0 \end{cases} \ee
is a state in which the observer's energy has a thermal distribution.    Time dependence is introduced in the usual way, by, for example,
$\h \a(t)=e^{\i H t}\h \a e^{-\i H t}$.

In the  limit that gravity is weakly coupled,
the de Sitter analogs of $F(t)$, $G(t)$ are supposed to be
\begin{align}\label{analog} \h F(t)& = \la\Psi_\max|\h\a(t)\h\b|\Psi_\max\ra=\la\Psi_\max|\a e^{-\i H t} e^{-\i p H } \Theta(q) e^{\i p H} \b|\Psi_\max\ra
\cr \h G(t)& =\la\Psi_\max|\h\b\h\a(t)|\Psi_\max\ra= \la\Psi_\max|\b e^{-\i p H} \Theta(q) e^{\i p H} e^{\i H t} \a|\Psi_\max\ra .    \end{align}
We have used the facts that $H\Psi_\max=0$, and  $\Theta\Psi_\max =\Psi_\max$.  
We observe now that
\be\label{observe}e^{-\i p H}\Theta(q) e^{\i p H}=\Theta(q-H)\ee
so  eqn. (\ref{analog}) simplifies to
\begin{align}\label{nanalog} \h F(t)& =\la\Psi_\max|\a e^{-\i H t}  \Theta(q-H) \b|\Psi_\max\ra
\cr \h G(t)& = \la\Psi_\max|\b  \Theta(q-H)e^{\i H t} \a|\Psi_\max\ra .    \end{align}
We expect $\h F(t)=\h G(t)$, since either of these functions is supposed to be $\Tr\, \h\a(t)\b$, where $\Tr $ is the trace of the Type II$_1$ algebra (as opposed
to the trace in the Hilbert space of an underlying thermal system, which has been denoted $\tr$ -- and which anyway does not really exist in the case of de Sitter space).

Now let us write a Fourier-transformed version of these last two equations.   In the formula for $\h F(t)$, we see that a contribution that varies with $t$ as $e^{-\i \omega t}$ comes
from intermediate states with $H=\omega$, but in the formula for $\h G(t)$,  such a contribution comes from states with $H=-\omega$.  The upshot of this
is that Fourier-transformed formulas can be written for $\h F$ and $\h G$ that are just analogous to those of eqn. (\ref{fourier}), but with an extra factor
involving the expectation value in the state $\Psi_\beta$ of $\Theta(q\mp \omega)$: 
\begin{align}\label{nourier} \h F(t) &=\int_{-\infty}^\infty \frac{\d \omega}{2\pi} e^{-\i \omega t} f(\omega)\la\Psi_\beta|\Theta(q-\omega)|\Psi_\beta\ra
\cr \h G(t)&=\int_{-\infty}^\infty \frac{\d \omega}{2\pi} e^{-\i \omega t} g(\omega) \la\Psi_\beta|\Theta(q+\omega)|\Psi_\beta\ra  .
\end{align}
Using the definition of $\Psi_\beta$, we can find explicit formulas for $u_\beta(\omega)= \la\Psi_\beta|\Theta(q-\omega)|\Psi_\beta\ra$ and
$v_\beta(\omega)=\la\Psi_\beta|\Theta(q+\omega)|\Psi_\beta\ra$:
\begin{align}\label{explicit} u_\beta(\omega)&=\begin{cases} e^{-\beta \omega} & \omega\geq 0 \cr 1& \omega<0\end{cases} \cr
                                           v_\beta(\omega)&=\begin{cases} 1 & \omega\geq 0 \cr e^{\beta \omega} & \omega<0.\end{cases} \end{align} 
 Note that
 \be\label{when}u_\beta(\omega)v_\beta(\omega)^{-1}=e^{-\beta\omega}.\ee                                          
We have then
\begin{align}\label{courier} \h F(t)&=\int_{-\infty}^\infty \frac{\d \omega}{2\pi} e^{-\i \omega t} \h f(\omega)
\cr \h G(t)&=\int_{-\infty}^\infty \frac{\d \omega}{2\pi} e^{-\i \omega t} \h g(\omega)
\end{align} 
with
\be\label{fourform} \h f(\omega) =u_\beta(\omega ) f(\omega), \hskip1cm \h g(\omega)= v_\beta(\omega) g(\omega) .     \ee
So using eqns. (\ref{undu}) and (\ref{when}), we have \be\label{goodform}
\h f(\omega) =u_\beta(\omega) e^{\beta\omega} g(\omega)= u_\beta(\omega) v_\beta(\omega)^{-1} e^{\beta\omega} \h g(\omega)=\h g(\omega),\ee
implying that $\h F(t)=\h G(t)$.
This confirms that the coupling to gravity has converted the thermal expectation value of an operator into a trace, and moreover that this trace is the expectation value
in the maximum entropy state $\Psi_\max$.

In sum, we have identified
a concrete example of including an observer in order to get a sensible answer in a cosmological model with a closed universe.  
 And, at least in the example of de Sitter space, we have understood  that gravity makes the notion of entropy better defined than it is in ordinary quantum field theory.  
 It is possible \cite{CLPW}  to probe more deeply  and show that entropy defined via the Type II$_1$ algebra agrees, up to an additive constant independent of the state,
 with the generalized gravitational entropy, as usually computed via quantum extremal surfaces.   It is also possible to give an analogous treatment of a black hole \cite{CLPW,Witt,CPW},
 involving in this case an algebra of Type II$_\infty$  and therefore no upper bound on the entropy.

\vskip1cm
 \noindent {\it {Acknowledgements}}  I thank  V. Hubeny, A. Jaffe, J. Kohn, H. Maxfield, R. Mazzeo,  G. Penington, and A. Strohmaier for comments and advice.
 Research supported in part by NSF Grant PHY-2207584.
 \bibliographystyle{unsrt}

\begin{thebibliography}{99}

\bibitem{CLPW}
V. Chandrasekharan, R. Longo, G. Penington, and E. Witten, ``An Algebra of Observables for De Sitter Space,'' arXiv:2206.10790.


\bibitem{Borch}
H. J. Borchers, ``Field Operators as ${\mathbb C}^\infty$ Functions In Spacelike Directions,'' Il Nuovo Cimento {\bf 33} (1964) 1.

\bibitem{Borch2}
H. J. Borchers, ``Uber die Vollst\"{a}ndigkeit lorentzinvarianter Felder in einer zeitartigen R\"{o}hre,''
Il Nuovo Cimento {\bf 19}  (1961) 787. 

\bibitem{Araki}
H. Araki, ``A Generalization Of Borchers' Theorem,'' Helv. Phys. Acta 36 (1963) 132-9.


\bibitem{Stroh}A. Strohmaier, ``On the Local Structure of the Klein-Gordon Field on Curved Spacetimes,'' Lett. Math. Phys.
{\bf 54} (2000) 249-61.

\bibitem{SW} A. Strohmaier and E. Witten, ``Analytic States in Quantum Field Theory on Curved Spacetimes,'' arXiv:2302.02709.

\bibitem{SW2}A. Strohmaier and E. WItten, ``The Timelike Tube Theorem in Curved Spacetime,''
arXiv:2303.16380.


\bibitem{Unruh}W. G. Unruh, ``Notes on Black Hole Evaporation,'' Phys. Rev. {\bf D14} (1976) 870.


\bibitem{Jaffe}
A. Jaffe, ``Wick Polynomials at a Fixed Time,'' J. Math. Phys. {\bf 7} (1966) 1250.

\bibitem{BF}
H. Bostelman and C. J. Fewster, ``Quantum Inequalities From Operator Product Expansions,'' Commun. Math. Phys. {\bf 292} (2009) 761-95,  arXiv:0812.4760.


\bibitem{Straube}
E. J. Straube, ``Harmonic and Analytic Functions Admitting a Distribution
Boundary Value,'' Annali della Scuola Normale Superiore di Pisa 4${e}$ Serie, {\bf 11} (1984) 559-91.


\bibitem{HW}
S. Hollands and R. M. Wald, ``Axiomatic Quantum Field Theory in Curved Spacetime,'' Commun. Math. Phys. {\bf 293} (2010) 85-125.

\bibitem{K}
M. Keyl, ``Quantum Fields on Timelike Curves,'' arXiv:math-ph/0012024.

\bibitem{Fewster}
C. J. Fewster, ``Lectures on Quantum Energy Inequalities,''
arXiv:1208.5399.





\bibitem{Smoller} J. Smoller, {\it Shock Waves and Reaction-Diffusion Equations}, second edition (Springer-Verlag, 2012).

\bibitem{Tataru} D. Tataru, ``Unique Continuation For Operators With Partially Analytic Coefficients,'' J. Math. Pures Appl. {\bf 78} (1999) 505-21.

\bibitem{casini1}H. Casini, M. Huerta, J. M. Magan, and D. Pontello, ``Entropic
Order Parameters for the Phases of QFT,'' JHEP {\bf 04} (2021) 277, arXiv:2008.11748.

\bibitem{casini2}H. Casini and J. M. Magan, ``On Completeness and Generalized Symmetries in Quantum Field Theory,'' arXiv:2110.11358.

\bibitem{Harlow}
D. Harlow, ``Wormholes, Emergent Gauge Fields, and the Weak Gravity Conjecture,'' JHEP {\bf 01} (2016) 122, arXiv:1510.07911

\bibitem{Polchinski}
J. Polchinski, ``Monopoles, Duality, and String Theory,'' Int. J. Mod. Phys. A {\bf 19S1} (2004) 145-56, hep-th/0304042.

\bibitem{BS}
T. Banks and N. Seiberg, ``Symmetries and Strings in Field Theory and Gravity,'' Phys. Rev. {\bf D83} (2011) 084019, arXiv:1011.5120.

\bibitem{HO}D. Harlow and H. Ooguri, ``Symmetries in Quantum Field Theory and Quantum Gravity,'' Commun. Math. Phys. {\bf 383}
 (2021) 3, 1669-1804, arXiv:1810.05338.
 
 \bibitem{GKSW} 
 D. Gaiotto, A. Kapustin, N. Seiberg, and B. Willett, ``Generalized Global Symmetries,'' JHEP {\bf 02} (2015) 172, arXiv:1412.5148.
 
  \bibitem{EW}
 E. Witten, ``Symmetry and Emergence,'' Nat. Phys. {\bf 14} (2018) 116-9, arXiv:1710.01791.
 

 
 
 \bibitem{BDHM}
 T. Banks, M. R. Douglas, G. T. Horowitz, and E. Martinec,
 ``AdS Dynamics From Conformal Field Theory,'' hep-th/9808016.
 
 \bibitem{B} I. Bena, ``On the Construction of Local Fields in the Bulk of AdS$_5$ and Other Spaces,'' Phys. Rev. {\bf D62} (2000) 066007, hep-th/9905186.
 
 \bibitem{Bal}
 V. Balasubramanian, P. Kraus and A.E. Lawrence, ``Bulk Versus Boundary Dynamics in
Anti-de Sitter Space-Time,'   Phys. Rev. { \bf D59} (1999) 046003, hep-th/9805171.
 
 \bibitem{HKLL} A. Hamilton, D. Kabat, G. Lifschytz, and D. A. Lowe,
 ``Holographic Representation of Local Bulk Operators,''  Phys.Rev. {\bf D74} (2006) 066009, 
 hep-th/0606141.
 
 \bibitem{HKLL2}
 A. Hamilton, D. Kabat, G. Lifschytz, and D. A. Lowe, ``Local Bulk Operators in AdS/CFT:
 A Holographic Description of the Black Hole Interior,'' Phys. Rev. {\bf D75} (2007) 106001, hep-th/0612053.
 
 \bibitem{KLL} 
 D. Kabat, G. Lifschytz, and D. A. Lowe, ``Constructing
 Local Bulk Observables in Interacting AdS/CFT,'' Phys. Rev. {\bf D83} (2011) 106009, arXiv:1102.2910.
 
 \bibitem{HMPS}I. Heemskerk, D. Marolf, J. Polchinski, and J. Sully, ``Bulk and Transhorizon
 Measurements in AdS/CFT,'' JHEP {\bf 10} (2012) 165,  arXiv:1201.3664.
 
 \bibitem{Mor} I. A. Morrison, ``Boundary-to-bulk Maps for AdS Causal Wedges and the
 Reeh-Schlieder Property in Holography,'' JHEP {\bf 05} (2014) 053,  arXiv:1403.3426.
 
 \bibitem{H}
 V. E. Hubeny, ``Covariant Residual Entropy,''   JHEP {\bf 09} (2014) 156, arXiv:1406.4611.
 
 \bibitem{Wald}R. M. Wald, {\it General Relativity} (University of Chicago, 1984).
 
 \bibitem{GaoWald}S. Gao and R. M. Wald, ``Theorems on Gravitational
 Time Delay and Related Issues,''
 Class. Quant. Grav. {\bf 17} (2000) 4999-5008, gr-qc/0007021.
 


 
 \bibitem{LightRays}E. Witten, ``Light Rays, Singularities, and All That,'' Rev. Mod. Phys.
 {\bf 92} (2020) 045004, arXiv:1901.03928.
 
 \bibitem{GH}G. W. Gibbons and S. W. Hawking, ``Cosmological Event Horizons,
 Thermodynamics, and Particle Creation,'' Phys. Rev. {\bf D15} (1977) 2738-51.
 
 \bibitem{FHN}
 R. Figari, R. Hoegh-Krohn, and C. R. Nappi, ``Interacting Relativistic Boson Fields in the
De Sitter Universe With Two Space-Time Dimensions,''  Commun. Math. Phys. {\bf 44} (1975)
265-278.
 



 \bibitem{Sewell}G. L. Sewell, ``Quantum Fields On Manifolds: PCT
 and Gravitationally Induced Thermal States,'' Ann. Phys. {\bf 141} (1982) 201-24.
 

\bibitem{Maeda} K. Maeda, T. Koike, M. Narita, A. Ishibashi, ``Upper Bound for Entropy in Asymptotically de Sitter Space-time,'' Phys. Rev. D {\bf 57(6)} (1998) 3503.


\bibitem{BoussoOne}R. Bousso, ``Positive Vacuum Energy and the $N$ Bound,'' JHEP {\bf 11} (2000) 038, hep-th/0012052.

\bibitem{BoussoTwo} R. Bousso, ``Bekenstein Bounds in de Sitter and Flat Space,'' JHEP {\bf 04} (2001) 035, hep-th/0010252.

\bibitem{Banks}T. Banks, ``Cosmological Breaking of Supersymmetry?  Or Little Lambda Goes Back to the Future, II,'' Int. J. Mod. Phys. {\bf A16}
(2001) 910-921,  hep-th/0007146.

\bibitem{BanksFischler}T. Banks and W. Fischler, ``$M$-Theory Observables For Cosmological Spacetimes,'' arXiv:hep-th/0102077.



\bibitem{BanksOne}T. Banks, ``More Thoughts on the Quantum Theory
of Stable de Sitter Space,'' arXiv:hep-th/0503066.

\bibitem{BanksTwo}T. Banks, B. Fiol, and A. Morisse, ``Towards a Quantum
Theory of de Sitter Space,'' arXiv:hep-th/0609062.

\bibitem{BFTwo}T. Banks and W. Fischler, ``The Holographic Model Of Cosmology,'' arXiv:1806.01749.

\bibitem{SusskindA}
L. Susskind, ``De Sitter Holography: Fluctuations, Anomalous Symmetry, and Wormholes,''  Universe {\bf 7} (2021) 464,
arXiv:2106.03964.

\bibitem{Susskind}
L. Susskind, ``Black Holes Hint Towards De Sitter-Matrix Theory,''
arXiv:2109.01322.

\bibitem{DST} X. Dong, E. Silverstein, and G. Torroba,
``De Sitter Holography and Entanglement Entropy,''
arXiv:1804.08623.

\bibitem{DF} P. Draper and S. Farkas,
``De Sitter Black Holes as Constrained States in the Euclidean Path Integral,''
Phys,. Rev. {\bf D105} (2022) 126022.

\bibitem{BD} T. Banks and P. Draper, ``Comments on the Entanglement Spectrum of
de Sitter Space,'' JHEP {\bf 01} (2023) 135, arXiv:2209.08991.

\bibitem{SB}
H. Lin and L. Susskind, ``Infinite Temperature's Not So Hot,'' arXiv:2206.01083.

 \bibitem{Takesaki}
M. Takesaki, ``Duality for Crossed Products and the Structure of von Neumann Algebras of Type III,'' Acta Mathematica 131 (1973) 249-310.


\bibitem{Sorce}
J. Sorce, ``Notes on the Type Classification of von Neumann Algebras,'' arXiv:2302.01958.

\bibitem{LL}
S. Leutheusser and H. Liu, ``Causal Connectability Between Quantum Systems and the Black Hole
Interior in Holographic Duality,'' arXiv:2110.05497.

\bibitem{LLtwo}S. Leutheusser and H. Liu, ``Emergent Times in Holographic Duality,'' 
arXiv:2112.12156.

\bibitem{Witt}
E. Witten, ``Gravity and the Crossed Product,''   JHEP {\bf 10} (2022) 008, arXiv:2112.12828.


 \bibitem{Segal}
 I. E. Segal, ``A Note on the Concept of Entropy,'' J. Math. Mech. {\bf 9} (1960) 623-9.
 
 \bibitem{LW} R. Longo and E. Witten, ``A Note On Continuous Entropy,'' arXiv:2202.03357.
 


\bibitem{CPW}
V. Chandraskharan, G. Penington, and E. Witten, ``Large $N$ Algebras and Generalized Entropy,''
arXiv:2206.10780.


\end{thebibliography}

\end{document}